\documentclass[11pt,letterpaper]{article}

\def\showauthornotes{1}
\def\showtableofcontents{1}
\def\showkeys{0}
\def\showdraftbox{0}
\def\showcolorlinks{1}
\def\usemicrotype{1}
\def\showfixme{1}

\usepackage{etex}

\usepackage[l2tabu, orthodox]{nag}

\usepackage{xspace,enumerate}

\usepackage[dvipsnames]{xcolor}

\usepackage[T1]{fontenc}
\usepackage[full]{textcomp}

\usepackage[american]{babel}

\usepackage{mathtools}

\usepackage{amsthm}

\newtheorem{theorem}{Theorem}[section]
\newtheorem*{theorem*}{Theorem}

\newtheorem{proposition}[theorem]{Proposition}
\newtheorem*{proposition*}{Proposition}
\newtheorem{lemma}[theorem]{Lemma}
\newtheorem*{lemma*}{Lemma}

\newtheorem*{corollary*}{Corollary}
\newtheorem*{conjecture*}{Conjecture}
\newtheorem{fact}[theorem]{Fact}
\newtheorem*{fact*}{Fact}

\newtheorem*{hypothesis*}{Hypothesis}
\newtheorem{conjecture}[theorem]{Conjecture}

\theoremstyle{definition}
\newtheorem{definition}[theorem]{Definition}

\theoremstyle{remark}
\newtheorem{claim}[theorem]{Claim}
\newtheorem*{claim*}{Claim}

\newtheorem*{remark*}{Remark}

\newtheorem*{observation*}{Observation}

\usepackage[
letterpaper,
top=1in,
bottom=1in,
left=1in,
right=1in]{geometry}

\usepackage{newpxtext} 
\usepackage{textcomp} 
\usepackage[varg,bigdelims]{newpxmath}
\usepackage[scr=rsfso]{mathalfa}
\usepackage{bm} 
\let\mathbb\varmathbb

\ifnum\showkeys=1
\usepackage[color]{showkeys}
\fi

\ifnum\showcolorlinks=1
\usepackage[
pagebackref,
colorlinks=true,
urlcolor=blue,
linkcolor=blue,
citecolor=OliveGreen,
]{hyperref}
\fi

\ifnum\showcolorlinks=0
\usepackage[
pagebackref,
colorlinks=false,
pdfborder={0 0 0}
]{hyperref}
\fi

\usepackage{prettyref}

\newcommand{\savehyperref}[2]{\texorpdfstring{\hyperref[#1]{#2}}{#2}}

\newrefformat{eq}{\savehyperref{#1}{\textup{(\ref*{#1})}}}
\newrefformat{eqn}{\savehyperref{#1}{\textup{(\ref*{#1})}}}
\newrefformat{lem}{\savehyperref{#1}{Lemma~\ref*{#1}}}
\newrefformat{def}{\savehyperref{#1}{Definition~\ref*{#1}}}
\newrefformat{thm}{\savehyperref{#1}{Theorem~\ref*{#1}}}
\newrefformat{prog}{\savehyperref{#1}{Program~\ref*{#1}}}
\newrefformat{cor}{\savehyperref{#1}{Corollary~\ref*{#1}}}
\newrefformat{cha}{\savehyperref{#1}{Chapter~\ref*{#1}}}
\newrefformat{sec}{\savehyperref{#1}{Section~\ref*{#1}}}
\newrefformat{app}{\savehyperref{#1}{Appendix~\ref*{#1}}}
\newrefformat{tab}{\savehyperref{#1}{Table~\ref*{#1}}}
\newrefformat{fig}{\savehyperref{#1}{Figure~\ref*{#1}}}
\newrefformat{hyp}{\savehyperref{#1}{Hypothesis~\ref*{#1}}}
\newrefformat{alg}{\savehyperref{#1}{Algorithm~\ref*{#1}}}
\newrefformat{rem}{\savehyperref{#1}{Remark~\ref*{#1}}}
\newrefformat{item}{\savehyperref{#1}{Item~\ref*{#1}}}
\newrefformat{step}{\savehyperref{#1}{step~\ref*{#1}}}
\newrefformat{conj}{\savehyperref{#1}{Conjecture~\ref*{#1}}}
\newrefformat{fact}{\savehyperref{#1}{Fact~\ref*{#1}}}
\newrefformat{prop}{\savehyperref{#1}{Proposition~\ref*{#1}}}
\newrefformat{prob}{\savehyperref{#1}{Problem~\ref*{#1}}}
\newrefformat{claim}{\savehyperref{#1}{Claim~\ref*{#1}}}
\newrefformat{relax}{\savehyperref{#1}{Relaxation~\ref*{#1}}}
\newrefformat{red}{\savehyperref{#1}{Reduction~\ref*{#1}}}
\newrefformat{part}{\savehyperref{#1}{Part~\ref*{#1}}}

\newcommand{\Sref}[1]{\hyperref[#1]{\S\ref*{#1}}}

\usepackage{nicefrac}

\ifnum\usemicrotype=1
\usepackage{microtype}
\fi

\ifnum\showauthornotes=1
\newcommand{\Authornote}[2]{{\sffamily\small\color{red}{[#1: #2]}}}
\newcommand{\Authornotecolored}[3]{{\sffamily\small\color{#1}{[#2: #3]}}}
\newcommand{\Authorcomment}[2]{{\sffamily\small\color{gray}{[#1: #2]}}}
\newcommand{\Authorstartcomment}[1]{\sffamily\small\color{gray}[#1: }

\newcommand{\Authorfnote}[2]{\footnote{\color{red}{#1: #2}}}
\newcommand{\Authorfixme}[1]{\Authornote{#1}{\textbf{??}}}
\newcommand{\Authormarginmark}[1]{\marginpar{\textcolor{red}{\fbox{\Large #1:!}}}}
\else
\newcommand{\Authornote}[2]{}
\newcommand{\Authornotecolored}[3]{}
\newcommand{\Authorcomment}[2]{}
\newcommand{\Authorstartcomment}[1]{}

\newcommand{\Authorfnote}[2]{}
\newcommand{\Authorfixme}[1]{}
\newcommand{\Authormarginmark}[1]{}
\fi

\definecolor{forestgreen(traditional)}{rgb}{0.0, 0.27, 0.13}

\ifnum\showfixme=0

\fi

\usepackage{boxedminipage}

\newcommand{\Paren}[1]{\left(#1\right)}

\newcommand{\Brac}[1]{\left[#1\right]}

\newcommand{\abs}[1]{\lvert#1\rvert}
\newcommand{\Abs}[1]{\left\lvert#1\right\rvert}

\newcommand{\norm}[1]{\lVert#1\rVert}
\newcommand{\Norm}[1]{\left\lVert#1\right\rVert}

\newcommand{\iprod}[1]{\langle#1\rangle}
\newcommand{\Iprod}[1]{\left\langle#1\right\rangle}

\newcommand{\Esymb}{\mathbb{E}}
\newcommand{\Psymb}{\mathbb{P}}
\newcommand{\Vsymb}{\mathbb{V}}

\DeclareMathOperator*{\E}{\Esymb}
\DeclareMathOperator*{\Var}{\Vsymb}
\DeclareMathOperator*{\ProbOp}{\Psymb}

\renewcommand{\Pr}{\ProbOp}

\newcommand{\tensor}{\otimes}

\newcommand{\textparen}[1]{\text{(#1)}}
\newcommand{\using}[1]{\textparen{using #1}}

\ifx\because\undefined
\newcommand{\because}[1]{\textparen{because #1}}
\else
\renewcommand{\because}[1]{\textparen{because #1}}
\fi

\newcommand{\sbits}{\{\pm1\}}

\newcommand{\mper}{\,.}
\newcommand{\mcom}{\,,}

\newcommand\bdot\bullet

\ifx\mathds\undefined 
\DeclareMathOperator{\Ind}{\mathbb{I}}
\else
\DeclareMathOperator{\Ind}{\mathds 1}}
\fi

\DeclareMathOperator{\Tr}{Tr}

\DeclareMathOperator{\argmax}{argmax}

\DeclareMathOperator{\sign}{sign}

\DeclareMathOperator{\rank}{rank}

\newcommand{\R}{\mathbb R}
\newcommand{\C}{\mathbb C}

\newcommand{\cK}{\mathcal K}

\newcommand{\cN}{\mathcal N}

\newcommand{\cS}{\mathcal S}

\renewcommand{\leq}{\leqslant}

\renewcommand{\geq}{\geqslant}

\ifnum\showdraftbox=1
\newcommand{\draftbox}{\begin{center}
  \fbox{%
    \begin{minipage}{2in}%
      \begin{center}%
          \Large\textsc{Working Draft}\\%
        Please do not distribute%
      \end{center}%
    \end{minipage}%
  }%
\end{center}
\vspace{0.2cm}}
\else
\newcommand{\draftbox}{}
\fi

\let\epsilon=\varepsilon

\numberwithin{equation}{section}

\newcommand\MYcurrentlabel{xxx}

\newcommand{\MYstore}[2]{%
  \global\expandafter \def \csname MYMEMORY #1 \endcsname{#2}%
}

\newcommand{\MYload}[1]{%
  \csname MYMEMORY #1 \endcsname%
}

\newcommand{\MYnewlabel}[1]{%
  \renewcommand\MYcurrentlabel{#1}%
  \MYoldlabel{#1}%
}

\newcommand{\MYdummylabel}[1]{}

\newcommand{\torestate}[1]{%
  \let\MYoldlabel\label%
  \let\label\MYnewlabel%
  #1%
  \MYstore{\MYcurrentlabel}{#1}%
  \let\label\MYoldlabel%
}

\newcommand{\restatetheorem}[1]{%
  \let\MYoldlabel\label
  \let\label\MYdummylabel
  \begin{theorem*}[Restatement of \prettyref{#1}]
    \MYload{#1}
  \end{theorem*}
  \let\label\MYoldlabel
}

\newcommand{\restatelemma}[1]{%
  \let\MYoldlabel\label
  \let\label\MYdummylabel
  \begin{lemma*}[Restatement of \prettyref{#1}]
    \MYload{#1}
  \end{lemma*}
  \let\label\MYoldlabel
}

\newcommand{\restateprop}[1]{%
  \let\MYoldlabel\label
  \let\label\MYdummylabel
  \begin{proposition*}[Restatement of \prettyref{#1}]
    \MYload{#1}
  \end{proposition*}
  \let\label\MYoldlabel
}

\newcommand{\restatefact}[1]{%
  \let\MYoldlabel\label
  \let\label\MYdummylabel
  \begin{fact*}[Restatement of \prettyref{#1}]
    \MYload{#1}
  \end{fact*}
  \let\label\MYoldlabel
}

\newcommand{\restate}[1]{%
  \let\MYoldlabel\label
  \let\label\MYdummylabel
  \MYload{#1}
  \let\label\MYoldlabel
}

\newcommand{\e}{\epsilon}

\let\origparagraph\paragraph
\renewcommand{\paragraph}[1]{\origparagraph{#1.}}

\allowdisplaybreaks

\usepackage{paralist}

\usepackage{comment}

\usepackage{graphicx}

\usepackage{relsize}
\usepackage[font=footnotesize]{caption}

\DeclareMathOperator{\Span}{span}

\DeclareMathOperator{\Id}{\mathrm{Id}}

\DeclareUrlCommand\email{}

\newcommand{\tr}{\textup{tr}}

\let\pref=\prettyref

\title{Matrix Discrepancy from Quantum Communication}

\author{Samuel B. Hopkins \\ UC Berkeley \and Prasad Raghavendra \\ UC Berkeley \and Abhishek Shetty \\ UC Berkeley}

\begin{document}

\maketitle
\draftbox

\begin{abstract}
We develop a novel connection between discrepancy minimization and (quantum) communication complexity.
As an application, we resolve a substantial special case of the \emph{Matrix Spencer} conjecture.
In particular, we show that for every collection of symmetric $n \times n$ matrices $A_1,\ldots,A_n$ with $\|A_i\| \leq 1$ and $\|A_i\|_F \leq n^{1/4}$ there exist signs $x \in \{ \pm 1\}^n$ such that the maximum eigenvalue of $\sum_{i \leq n} x_i A_i$ is at most $O(\sqrt n)$.
We give a polynomial-time algorithm based on partial coloring and semidefinite programming to find such $x$.

Our techniques open a new avenue to use tools from communication complexity and information theory to study discrepancy.
The proof of our main result combines a simple compression scheme for transcripts of repeated (quantum) communication protocols with quantum state purification, the Holevo bound from quantum information, and tools from sketching and dimensionality reduction.
Our approach also offers a promising avenue to resolve the Matrix Spencer  conjecture completely -- we show it is implied by a natural conjecture in quantum communication complexity.

\end{abstract}

\ifnum\showtableofcontents=1
{
\tableofcontents
\thispagestyle{empty}
 }
\fi

\newpage

\setcounter{page}{1}

\section{Introduction}

In this paper we study \emph{discrepancy minimization} for matrices.
To set up our main problem, let us begin with the classic result of Spencer, ``six standard deviations suffice.''
Let $v_1,\ldots,v_n \in \R^m$ have $\|v_i\|_\infty \leq 1$.
The goal is to assign signs $x_1,\ldots,x_n \in \{ \pm 1\}$ to the vectors so as to minimize $\|\sum_{i \leq n} x_i v_i\|_\infty$.
As a shorthand, we often call the latter quantity the discrepancy of $x$.
For some intuition, note that if the vectors $v_i \in \{0,1\}^m$ then, treating them as incidence vectors, they define a set system with $n$ atoms and $m$ subsets.
The goal then becomes to assign $x_1,\ldots,x_n$ so as to minimize the maximum difference between the number of $+1$'s and $-1$'s in each set.

Choosing $x_1,\ldots,x_n$ at random presents a natural benchmark -- in this case, $\E_{x \sim \{ \pm 1\}^n} \|\sum x_i v_i \|_\infty \leq O(\sqrt{n \log m})$, by a Chernoff/union bound argument.
While many similar applications of the union bound in combinatorics give tight results, Spencer's result remarkably shows that for any $v_1,\ldots,v_n$, this bound can in fact be beaten.

\begin{theorem}[\cite{spencer1985six}]
    For all $v_1,\ldots,v_n \in \R^n$ with $\|v_i\|_\infty \leq 1$, there exist $x_1,\ldots,x_n \in \left\{ -1,1 \right\} $ such that $\| \sum_{i \leq n} x_i v_i \|_\infty \leq O(\sqrt {n \log(m/n)})$.
\end{theorem}

In particular, if $m = O(n)$, Spencer's result shows that a signing of discrepancy $O(\sqrt n)$ always exists.\footnote{And, in fact, the constant in the big-$O$ is at most $6$, hence the name.} Spencer's original result was nonconstructive, but a following a breakthrough by Bansal \cite{bansal2010constructive}, several polynomial-time algorithms are now known to find such a signing, e.g. \cite{lovettmeka,rothvossconvex,eldan2018efficient}.

\paragraph{Matrix Discrepancy}
We generalize the preceding setting by replacing the vectors $v_1,\ldots,v_n$ with symmetric matrices $A_1,\ldots,A_n \in \R^{m \times m}$ having spectral norms $\|A_i\| \leq 1$.\footnote{We expect that the main results in this paper continue to hold if $\R$ is replaced by $\C$. Furthermore, if the matrices $A_i$ are not symmetric/Hermitian, they can be replaced by their ``Hermition dilations'' $\left ( \begin{matrix} 0 & A_i \\ A_i^\top & 0 \end{matrix} \right )$ without changing any of the asymptotic bounds in this paper.}
Now the goal is to find $x_1,\ldots,x_n$ to minimize the spectral norm $\|\sum_{i \leq n} x_i A_i\|$.
Note that we can recover the vector case by taking the $A_i$s to be diagonal, or more generally, commuting.

The \emph{matrix Chernoff} bound of Ahlswede and Winter shows that, as in the vector setting, randomly choosing $x$ gives a signing of discrepancy $\|\sum_{i \leq n} x_i A_i \| \leq O(\sqrt{n \log m})$ \cite{ahlswede2002strong}.
This inequality and its generalizations have become crucial tools in mathematics and theoretical computer science, including in applications of the probabilistic method, for instance in spectral graph theory and unsupervised learning, e.g. \cite{spielman2011graph,gross2011recovering}.
It is a natural question to ask whether it, too, can be improved by careful choice of $x$ -- this is the content of the \emph{Matrix Spencer conjecture}:

\begin{conjecture}[Matrix Spencer \cite{mekablog,zouzias2012matrix}]
    For all $A_1,\ldots,A_n \in \R^{m \times m}$ with $\|A_i\| \leq 1$ there exists $x \in \{ \pm 1\}^n $ such that $\| \sum_{i \leq n} x_i A_i \| \leq O(\sqrt{n \log(m/n)}$).
\end{conjecture}

Despite significant effort, this conjecture has remained largely open for a decade, with partial progress in the block-diagonal and rank-one cases \cite{levy2017deterministic,marcus2015interlacing,kyng2020four}.
Thus, resolving Matrix Spencer (even in a substantial special case) seems likely to lead to new techniques in discrepancy.

\subsection{Results}
We resolve the Matrix Spencer conjecture in the case that $A_1,\ldots,A_n$ have \emph{moderate} rank.
More formally, in addition to the assumption $\|A_i\| \leq 1$, we additionally assume that $\|A_i\|_F \leq n^{1/4}$, where $\|\cdot \|_F$ is the Frobenius norm.

\begin{theorem}[Moderate-Rank Matrix Spencer]
    \label{thm:matrix-spencer-intro}
    Let $A_1,\ldots,A_n \in \R^{m \times m}$ have $\|A_i\| \leq 1$ and $\|A_i\|_F \leq n^{1/4}$.
    Then there exists $x \in \{ \pm 1 \}^n$ such that $\|\sum_{i \leq n} x_i A_i \| \leq O(\sqrt{ n \log (m/n)})$.
    Furthermore, such an $x$ can be found in polynomial time.
\end{theorem}

Even in the presence of the ``moderate rank'' assumption $\|A_i\|_F \leq n^{1/4}$, our result captures settings where the looser bound $O(\sqrt{ n \log m})$ is un-improvable for randomly-chosen $x$ -- for instance, if the $A_i$'s are all diagonal with nonzero entries in the first $\sqrt n$ diagonal entries.\footnote{We thank Tselil Schramm and Boaz Barak for pointing this out.}
Thus, our result captures a novel improvement over the matrix Chernoff bound.

To prove Theorem~\ref{thm:matrix-spencer-intro}, we introduce a new approach to discrepancy minimization using (one-way) communication complexity.
In the matrix case, this connection leads us to quantum communication.
For starters, we give a new proof of Spencer's theorem: after translation into a (classical) communication problem, Spencer's theorem can be proved using a simple compression scheme for repeated communication protocols.
To prove our moderate-rank Matrix Spencer theorem, we combine a quantum analogue of this compression scheme with several other tools, including quantum state purification, sketching/dimensionality reduction, and consequences of the \emph{Holevo bound} from quantum information theory.

Discrepancy bounds proved using our techniques are automatically algorithmic.
In the vector (Spencer) case, our arguments give a new analysis of the randomized linear programming approach first analyzed by Eldan and Singh \cite{eldan2018efficient}.
In the matrix case, we give an analogous algorithm based on semidefinite programming.
(This algorithm uses a very different semidefinite program than Bansal's original use of semidefinite programming in the vector case.)

Without the ``moderate-rank'' assumption $\|A_i\|_F \leq n^{1/4}$, the $O(\sqrt n)$ discrepancy bound is tight.
This is witnessed by examples from the vector setting (in particular, rows of Hadamard matrices), meaning that the $O(\sqrt n)$ bound would be tight even for diagonal matrices $A_i$.
However, it remains open to determine if the $O(\sqrt n)$ bound is tight with the additional moderate-rank assumption.
Note that it cannot be tight for diagonal matrices under this assumption, since $n$ vectors in $\R^n$ with $\ell_2$ norms $n^{1/4}$ have discrepancy at most $\tilde{O}(n^{1/4})$ (the \emph{Koml\'os} setting) \cite{banaszczyk1998balancing}.
This suggests a number of interesting questions beyond Matrix Spencer: are there matrix analogues of other discrepancy bounds for vectors, for instance under $\ell_2$ assumptions (as in the Koml\'os setting) or $\ell_2$ and $\ell_\infty$ assumptions (like the Beck-Fiala setting)?

We hope that opening the way to use communication complexity techniques to prove results in discrepancy leads to future progress.
As an illustration, we show that our techniques offer a promising avenue to fully resolve the Matrix Spencer conjecture -- we now describe a natural conjecture in quantum communication complexity which would imply it.

To describe the conjecture we need a small amount of notation.
Let $\text{index} \, : \, \{-1,1 \}^n \times [n] \rightarrow \{-1,1\}$ be the \emph{index function}, given by $\text{index}(x,i) = x_i$.
The index function induces the following one-way communication problem between two players, Alice and Bob.
Alice receives $x \in \{ -1,1\}^n$ and Bob receives $i \in [n]$.
Alice sends Bob a message $a$, after which Bob must output a bit $b(a,i) \in \{-1,1\}$; their goal is to jointly compute $\text{index}(x,i)$.

The main question in one-way communication complexity is: how long must Alice's message be?
This could depend on several things:
\begin{itemize}
    \item The nature of Alice's message -- classical or quantum.
    \item The probability of success $\Pr(b(a,i) = x_i)$ (where the probability is over randomness in the protocol).
    \item The distribution of Alice and Bob's inputs -- they could be uniformly random, worst-case, or something else.
\end{itemize}
Later, we will thoroughly discuss the one-way communication complexity of the index function, after which the following conjecture will be less mysterious.
For now, we state the conjecture as an illustration of the surprising connection between discrepancy and communication.

\begin{conjecture}[Quantum One-Way Communication Complexity in the Small-Advantage Regime]
    \label{conj:quantum-one-way}
    Suppose Alice's message $\rho$ consists of $q$ qubits, and Bob has $\e$ advantage over random guessing in computing $x_i$ for a large set of indices $i$, in the following sense.
    For each $x$ there is a set of coordinates $S_x \subseteq [n]$ with $|S_x| \geq (1-\delta)n$ such that $\E_{x \sim \{ -1,1 \}^n} \min_{i \in S_x} \Pr(b(\rho,i) = x_i) \geq 1/2 + \e$.
    Then for every small-enough $\delta > 0$, if $\e \gg 1/\sqrt{n}$, Alice must send $q \geq \log(1/\e^2) + \Omega(\e^2 n)$ qubits.
\end{conjecture}

Note that Conjecture~\ref{conj:quantum-one-way} remains interesting even if $\delta = 0$; indeed, this special case is most interesting from a quantum communication point of view, and we expect that it already contains most of the challenge in proving the conjecture.

Using the same argument as for Theorem~\ref{thm:matrix-spencer-intro} but substituting the communication lower bound in Conjecture~\ref{conj:quantum-one-way} for a weaker version we prove in the course of proving Theorem~\ref{thm:matrix-spencer-intro}, our techniques show:

\begin{theorem}
    \label{thm:full-spencer-intro}
    Suppose Conjecture~\ref{conj:quantum-one-way} is true.
    Then the Matrix Spencer conjecture holds, and there is a polynomial-time algorithm based on semidefinite programming to find the signing it promises.
\end{theorem}

The classical analogue of Conjecture~\ref{conj:quantum-one-way} is true; we record a proof in this paper, although we believe it is probably known implicitly in the literature.
In fact, using our techniques, the classical analogue gives a new algorithmic proof of Spencer's theorem.
Our proof of Theorem~\ref{thm:matrix-spencer-intro} establishes a special case of Conjecture~\ref{conj:quantum-one-way} where Alice must send a \emph{pure state}, from which (with some work) we are able to deduce our moderate-rank Matrix Spencer theorem.

\subsection{Techniques}

\subsubsection{From Discrepancy to Communication}

\paragraph{Discrepancy Is Exactly Average-Bob One-Way Communication Complexity}
To build intuition, we start with the following simple observation.
Let $R_{\text{worst,unif}} = R_{\text{worst,unif}}(n,\e)$ be the minimum length of a message $a$ that Alice must send to Bob in a one-way classical protocol for the $n$-bit index function in order to achieve
\[
    \min_{x \in \{ \pm 1\}^n} \E_{i \sim [n]} \Pr(b(a,i) = x_i) \geq \frac 12 + \e\mper
\]
Here, the subscript ``worst,unif'' denotes that Alice's input is worst-case over $x \in \{ \pm 1\}^n$ and Bob's is uniform in $[n]$.
Similarly, define $Q_{\text{worst,unif}}$ for quantum one-way communication.
The following claim shows that \emph{lower bounds} on $R_{\text{worst,unif}}$ and $Q_{\text{worst,unif}}$ imply \emph{upper bounds} on discrepancy for vectors and matrices.

\begin{claim}
    For every integer $m > 0$, if $R_{\text{worst,unif}}(n,\e) > \log m + O(1)$ then for $v_1,\ldots,v_n \in \R^m$ with $\|v_i\|_\infty \leq 1$ there exists $x \in \{ -1,1 \}^n$ such that $\|\sum_{i \leq n} x_i v_i\|_\infty \leq 2 \e n$.
    Conversely, if $R_{\text{worst,unif}}(n,\e) < \log m - O(1)$, then there exist $v_1,\ldots,v_n \in \R^m$ with $\|v_i\|_\infty \leq 1$ and $x \in \{ \pm 1\}^n$ such that $\|\sum_{i \leq n} x_i v_i\|_\infty \leq 2\e n$.
    Furthermore, the same holds if we replace $R$ by $Q$ and the $v$s by matrices $A_1,\ldots,A_m \in \R^{m \times m}$ with $\|A_i\| \leq 1$.
\end{claim}
\begin{proof}
    We show one direction of the proof in the classical case; the other direction and the quantum case are similar.
    Suppose $v_1,\ldots,v_n \in \R^m$ have $\|v_i\|_\infty \leq 1$ but for every $x \in \{ \pm 1\}^n$ we have $\|\sum_{i \leq n} x_i v_i \|_\infty > 2 \e n$.
    Then for each $x$ we may associate a standard basis vector $y_x \in \R^m$ such that $|\iprod{y_x, \sum_{i \leq n} x_i v_i}| > 2 \e n$.
    This induces a $\log m + O(1)$ bit communication protocol as follows.
    On input $x$, Alice sends Bob the name of the coordinate $j$ represented by $y_x$, as well as the sign $s$ of $\iprod{y_x, \sum_{i \leq n} x_i v_i}$.
    Bob outputs a biased random bit $b(j,i)$ with expectation $s \cdot v_i(j)$.
    Then for each $x$ we can compute:
    \[
        \E_i\Brac{ \Pr(b(j,i) = x_i) } = \E_i \Brac{ \frac 12 + s \cdot x_i \cdot \frac {v_i(j)}{2}} > \frac 12 + \e\mper
    \]
    For the other direction, observe that a protocol for the index function where there are $m$ possible messages Alice may send induces a set of vectors $v_1,\ldots,v_n \in \{ \pm 1\}^m$ by writing out Bob's outputs.
    The success probability of this protocol gives a lower bound on the discrepancy of $v_1,\ldots,v_n$.
\end{proof}

In spite of its simplicity, we do not know how to use this connection between discrepancy and $R_{\text{worst,unif}}$ and $Q_{\text{worst,unif}}$ to prove any interesting discrepancy upper bounds.
The difficulty is that in the regime of interest for (Matrix) Spencer-type theorems, $m \approx n$ and $\e \approx 1/\sqrt{n}$.
That is, Bob has very tiny advantage over oblivious random guessing in determining $\text{index}(x,i)$, and Alice is sending just a logarithmic number of (qu)bits.
We are not aware of any direct techniques to lower bound $R_{\text{worst,unif}}$, let alone $Q_{\text{worst,unif}}$ in this regime.
(Of course, an indirect argument is available for $R_{\text{worst,unif}}$ by appealing to Spencer's discrepancy result.)

\paragraph{Trading Average Bob for Average Alice}
Our first key technical contribution is another connection between discrepancy and communication, but for $R_{\text{unif,worst}}$ and $Q_{\text{unif,worst}}$ rather than $R_{\text{worst,unif}},Q_{\text{worst,unif}}$ -- that is, now Alice's input will be random, but Bob's will be worst-case.
While this difference may seem small, the requirement that (for typical $x$) Bob has nontrivial advantage over random guessing in computing $x_i$ for all $i$ makes it much easier to prove lower bounds -- we will see why momentarily.
(Actually, our lower bounds will apply even when Bob has nontrivial advantage for, say, $0.9n$ coordinates $i \in [n]$ -- this technical improvement is important for the connection to discrepancy, but we will mainly ignore it for simplicity in this introduction.)

We now discuss the key lemma we prove connecting communication and \emph{partial coloring}, starting with the following standard definition:

\begin{definition}[Partial coloring]
    A partial coloring of matrices $A_1,\ldots,A_n$ with discrepancy $\Delta > 0$ is a vector $x \in [-1,1]^n$ such that $|x_i| = 1$ for a constant fraction of coordinates $i \in [n]$, and $\|\sum_{i \leq n} x_i A_i\| \leq \Delta$.\footnote{This is often called a ``fractional'' partial coloring in the literature; since all partial colorings in this paper are fractional we drop the modifier.}
(A similar definition applies for the case of vectors $v_1,\ldots,v_n$.)
\end{definition}

It is a standard result that Spencer-style discrepancy theorems can be proved by alternately finding partial colorings and removing vectors/matrices which have been fully colored (i.e., they have $|x_i| = 1$), so it suffices to prove the existence of partial colorings with small discrepancy.

For simplicity in this introduction, we restrict attention to the setting where the number of vectors/matrices is the same as the dimension -- i.e. $A_1,\ldots,A_n \in \R^{n \times n}$ or $v_1,\ldots,v_n \in \R^n$ -- in which case we are looking for partial colorings of discrepancy $O(\sqrt n)$.
And, for now, we drop algorithmic considerations and worry only about the existence of partial colorings.

\begin{lemma}[Special case of the Compress or Color Lemma (Lemma~\ref{lem:compress-or-color}), informal]
    \label{lem:compress-or-color-intro}
    Suppose $A_1,\ldots,A_n \in \R^{n \times n}$ with $\|A_i\| \leq 1$ lack a partial coloring with discrepancy $O(\sqrt n)$.
    Then there is a quantum one-way communication protocol for the $n$-bit index function of the following form. 
    Alice sends a $\log n + 1$ qubit message $\rho$. 
    If Bob gets input $i$, he measures $\rho$ in the eigenbasis of the matrix $\left ( \begin{matrix} A_i & 0 \\ 0 & - A_i \end{matrix} \right )$, receiving an eigenvalue $\lambda_i$ as an outcome; then he outputs a random bit $b(\rho,i)$ with bias $\lambda_i$.
    This protocol has the following guarantee: for every $x \in \{\pm 1\}^n$ there is a set $S_x \subseteq [n]$ with $|S_x| \geq 0.99n$ such that
    \[
        \E_{x \in \{ \pm 1 \}^n} \min_{i \in S_x} \Pr(b(\rho,i) = x_i) = \frac 12 + \e, \qquad \e \gg \frac 1 {\sqrt n}\mper
    \]
\end{lemma}

From this lemma we can see the origin of Conjecture~\ref{conj:quantum-one-way} and Theorem~\ref{thm:full-spencer-intro}.
It also shows that to prove our moderate-rank Matrix Spencer theorem, it suffices to rule out $\log n$-qubit protocols for the index function with advantage $\e \gg 1/\sqrt n$ where Bob's measurement matrices have $\|A_i\|_F \leq n^{1/4}$.

We make a few more remarks about the Compress or Color Lemma (Lemma~\ref{lem:compress-or-color}) before we move on to communication lower bounds, since we think the general version of the lemma is of independent interest.

\subparagraph{General norms} In full generality, the lemma says that for \emph{any} collection of vectors $v_1,\ldots,v_n$ and \emph{any} norm $\|\cdot \|$, either $v_1,\ldots,v_n$ admit a small-$\|\cdot\|$-discrepancy partial fractional coloring (i.e. a coloring where $\|\sum_{i \leq n} x_i v_i\|$ is small) or $v_1,\ldots,v_n$ induce a certain kind of compression of the hypercube $\{ \pm 1\}^n$ into the dual ball of $\|\cdot \|$.
In the case that $\|\cdot\|$ is $\ell_\infty$, this compression turns out to be a classical communication protocol.
When $\|\cdot\|$ is the spectral norm, the result is a quantum communication protocol.

\subparagraph{Rademacher width} Second, the proof of the lemma goes via studying the \emph{Rademacher width} of the set of partial colorings.
The Rademacher width of a set $\cK \subseteq \R^n$ is $\E_{g \sim \{ \pm 1\}^n} \max_{x \in \cK} \iprod{g,x}$ -- it is a standard measure of the size of $\cK$.
A long-established technique in discrepancy is to study the \emph{Gaussian volume} of the partial colorings.
By studying width instead, we can prove the lemma using tools from convex programming, in particular strong duality.
Gaussian width was previously studied in the context of discrepancy by Eldan and Singh and by Reis and Rothvoss \cite{eldan2018efficient,reis2020linear}; we borrow some tools from Eldan and Singh in the proof of the Compress or Color Lemma.
(The switch from Gaussian to Rademacher width -- that is, using $\pm 1$-valued coordinates in $g$ -- is just a technical convenience.)

\subparagraph{Polynomial-time consequences} Finally, since the heart of the Compress or Color Lemma is a convex program, it also has algorithmic consequences when that convex program is efficiently solvable.
In particular, our proof of the contrapositive of the above statement, that communication lower bounds imply the existence of partial colorings, actually proves something stronger: such a partial coloring can be found (with high probability) by drawing a random $g \sim \{ \pm 1\}^n$ and maximizing $\iprod{g,x}$ over partial colorings $x$ with low discrepancy.
Note that this is a convex program -- in particular, for the matrix discrepancy setting, it is a semidefinite program.

\subsubsection{Communication Lower Bounds in the Small-Advantage Regime}
Now that we have seen that communication lower bounds imply the existence of partial colorings, we need to prove some communication lower bounds.

\paragraph{Classical}
To build some intuition, we start with the classical case.
According to Lemma~\ref{lem:compress-or-color-intro} (instantiated with diagonal matrices), to prove that a partial coloring of any $v_1,\ldots,v_n \in \R^n$ with $\|v_i\|_\infty \leq O(\sqrt n)$ exists it will suffice to rule out $\log n$-bit one-way protocols for the $n$-bit index function where Alice's input is random, Bob's is worst case, and they have advantage $\e \gg 1/\sqrt n$ over oblivious random guessing.
(To avoid technicalities, for now we consider protocols where Bob has this advantage on all inputs $i \in [n]$, rather than just $0.99n$ of them.)

To see the subtlety of the lower bound we need to establish, let us first consider what we could get from naive information-theoretic arguments.
By directly analyzing the mutual information between Alice's input and Bob's output, we could show that Alice must send at least $(1-H(1/2 + \e))n$ bits, where $H$ is the binary entropy function.
For small $\e$, we have $(1-H(1/2 + \e)) n \approx \e^2 n$ -- this lower bound degrades to just $O(1)$ when $\e \approx 1/\sqrt n$, while we need a bound larger than $\log n$.

Indeed, if Bob's input is also random, there is actually an $O(1)$-bit protocol achieving advantage $\e \gg 1/\sqrt n$.
Even with worst-case inputs, there is a $\log n + O(1)$-bit protocol based on Hadamard matrices which achieves advantage $\e \geq \Omega(1/\sqrt n)$.
(For both protocols, see Section~\ref{sec:tightness}.)
This shows that we must use worst-case-ness of Bob's input in our lower bound, and even when we do, our argument must be tight up to additive constants.
We now sketch a simple argument satisfying both of these requirements.

\begin{lemma}
    \label{lem:classical-lb}
  Any classical protocol for the $n$-bit index function achieving advantage $\e \gg 1/\sqrt n$ when Alice's input is uniformly random and Bob's is worst-case requires Alice to send more than $\log n$ bits.
    That is, $R_{\text{unif,worst}}(n,\e) > \log n$ when $\e \gg 1/\sqrt{n}$.
\end{lemma}
\begin{proof}[Proof sketch]
    Suppose for contradiction that a $\log n$ bit protocol exists with advantage $\e \gg 1/\sqrt n$.
    By repeating the protocol $O(1/\e^2) \ll n$ times, the players can amplify their success probability to $0.9$.
    Concretely, in this \emph{amplified protocol}, Alice receives $x$ and makes $O(1/\e^2)$ independent draws from the distribution over messages she would send on input $x$ in the original protocol.
    She sends all of these messages to Bob, who computes all of the outputs he would compute in the original protocol and takes a majority vote.
    (Note that this amplification relies on Bob having a worst-case input -- otherwise, Bob might already have success probability $0.9$ on a few inputs and exactly $1/2$ on the rest, in which case the amplification does not have the desired effect.)

    Naively, Alice is now sending $\Omega(\log n / \e^2) \gg n$ bits, but we claim that her message in the amplified protocol can be compressed down to $n/10$ bits (for appropriate $\e \gg 1/\sqrt n$).
    This leads to a contradiction, since she is sending Bob at least $n/2$ bits of information.

    To see this, observe that it actually suffices for Alice to send a histogram of her $O(1/\e^2)$ messages from the original-protocol distribution, since Bob does not need to know the ordering of the $O(1/\e^2)$ messages.
    Since Alice's individual messages are $\log n$ bits, there are only $n$ possible messages in the original protocol, so she is sending multi-subset/histogram of $[n]$ of size $O(1/\e^2)$.
    A simple counting argument shows that there are approximately $\binom{n}{O(1/\e^2)}$ such histograms (this is exactly the number if there are no repeated messages, but repeated messages do not change the asymptotics).
    Since $1/\e^2 \ll n$, Alice can now send just $\log \binom{n}{O(1/\e^2)} \ll n$ bits.
\end{proof}

Carrying out this argument carefully actually shows the following quantitative bounds, which may be independently interesting:
\begin{itemize}
    \item If $\e \gg 1/\sqrt n$, then $R_{\text{unif,worst}}(n,\e) \geq \log (1/\e^2) + \Omega(\e^2 n)$.
    \item If $\e \ll 1/\sqrt n$, then $R_{\text{unif,worst}}(n,\e) \geq \log n - \log \log (1/\e^2) - O(1)$.
\end{itemize}
This means that even for very small $\e$, like $2^{-n^{0.99}}$, Alice still must send $\Omega(\log n)$ bits.
Using our discrepancy-to-communication technique, this ``microscopic $\e$'' bound implies discrepancy bounds for set systems with many more atoms than sets -- with $n$ atoms and $m$ sets, discrepancy $O(\sqrt{m})$ is achievable.
(There are also generic reductions from $n \gg m$ to $n = m$, using linear programming.)

\paragraph{Quantum}

We turn to the case of quantum communication lower bounds, which we need for the matrix discrepancy setting.
To prove the (unrestricted) matrix Spencer conjecture, we would like to prove a quantum analogue of Lemma~\ref{lem:classical-lb}.
Unfortunately, the histogram-based compression used in the argument above seems inherently classical, so another idea is needed.

To prove moderate-rank matrix Spencer, our second key technical contribution is a quantum analogue of Lemma~\ref{lem:classical-lb} when Bob's measurement matrices have $\|A_i\|_F \leq n^{1/4}$.
By Lemma~\ref{lem:compress-or-color-intro}, this shows that partial colorings exist for every family of $A_1,\ldots,A_n \in \R^{n \times n}$ with $\|A_i\| \leq 1$ and $\|A_i\|_F \leq n^{1/4}$.

For simplicity in this introduction, consider the case that each matrix $A_i$ has all eigenvalues in $\{-1,0,1\}$, with at most $\sqrt n$ nonzero eigenvalues.
Alice gets a randomly chosen $x \in \{ \pm 1\}^n$ and sends Bob a $\log n$-qubit mixed state, represented by a density matrix $\rho = \rho_x$.
Given any input $i$, Bob measures $\rho_x$ with $A_i$, getting back an eigenvalue.
If he receives $-1$ or $1$, he outputs the result; otherwise he outputs $-1$ or $1$ uniformly at random.
We want to show:

\begin{lemma}
    \label{lem:quantum-lb}
  Alice and Bob cannot achieve success probability $1/2 + \e$ for $\e \gg 1/\sqrt n$ by the above protocol.
\end{lemma}

We now sketch the proof of Lemma~\ref{lem:quantum-lb}, which takes several ingredients.

\subparagraph{Ruling out pure-state protocols} The first step is to prove a lower bound against \emph{pure state} protocols with (potentially) \emph{full-rank measurements}.
That is, we consider the case that Alice actually sends a pure state $a_x \in \R^n$, and Bob is allowed to use any measurements $M_i \in \R^{n \times n}$ with $\pm 1$ eigenvalues.
In this case, we can use a similar amplify-then-compress approach as in the classical case.
We sketch the proof here -- for details, see Theorem~\ref{thm:rac-lb} for the communication lower bound and Lemma~\ref{lem:rac-lb-matrix-dependent} for a version adapted to the discrepancy bound we need to prove.

In a little bit more detail, to amplify from success probability $1/2 + \e$ to $0.9$, Alice sends $O(1/\e^2)$ copies of her message; Bob makes his measurement independently on each of them and takes a majority vote.
Naively, this requires Alice to send around $\log n /\e^2 \gg n$ qubits, but the state that Alice sends, $a_x^{\otimes O(1/\e^2)}$, actually lies in the \emph{symmetric subspace} of $(\R^n)^{\otimes O(1/\e^2)}$.
This subspace, $\Span \{ a^{\otimes O(1/\e^2)} \, : \, a \in \R^n \}$, has dimension roughly $\binom{n}{O(1/\e^2)}$ by the same counting argument as we used in the classical case, which means that Alice can compress her message into $\log \binom{n}{O(1/\e^2)} \ll n$ qubits.
As in the classical case, this argument crucially uses that Bob succeeds on any (worst-case) input $i \in [n]$.

Since Alice is sending a quantum state, this situation is no longer ruled out by classical information theory.
However, a known consequence of the Holevo bound from quantum information says that Alice cannot communicate the $n$ classical bits $x$ without sending $\Omega(n)$ qubits.
This is not a trivial consequence of the Holevo bound, since Bob cannot necessarily read more than one of the bits of $x$ without collapsing the state he is sent in a way which prevents reading any of the remaining bits.
This situation has been considered before, however: a result of Ambainis, Nayak, Ta-Shma, and Vazirani on \emph{quantum random access codes} shows that it is still impossible for Alice to send $\ll n$ qubits even if Bob can only read one coordinate of $x$ (so long as he may choose this coordinate at will) \cite{ambainis2002dense}.

\subparagraph{Reduction from moderate-rank to pure-state protocols}
We now sketch an argument that if there is a protocol of the type described in Lemma~\ref{lem:quantum-lb}, where Alice may be sending a mixed state $\rho_x$, then there is also a pure-state protocol of the sort we just ruled out.
We call this the \emph{purify-then-sketch} transformation (Lemma~\ref{lem:purify-then-sketch}).

First, we may assume that Alice's mixed states $\rho_x \in \R^{n \times n}$ have rank at most $\sqrt n$ -- this is because we can take them to be extremal solutions to a semidefinite program involving $n$ linear constraints, one for each matrix $A_i$ \cite{barvinok1995problems,pataki1998rank}.
Using this bound on the rank of the $\rho_x$'s, we can use quantum state purification to replace them with pure states $\tilde{a}_x$ of dimension $\R^{n^{3/2}}$; when Bob measures the purified states he uses measurement matrices $A_i \tensor I$, where $I$ is identity in $\sqrt n$ dimensions.
This gives a pure-state protocol with the same success probability as the protocol we started with (since the outcomes of Bob's measurements will have exactly the same distributions as before), but now Alice has to send $\tfrac 32 \log n$ qubits, so we cannot apply the above lower bound against pure-state protocols.

To fix this, Alice replaces $\tilde{a}_x$ with a random sketch $a_x = S \tilde{a}_x$ of it down to $n$ dimensions, and hence $\log n$ qubits.
(Here $S$ is a random sketching matrix).
Bob replaces $A_i \tensor I$ with $S(A_i \tensor I)S^\top$.
We argue that \emph{so long as $A_i$ has at most $\sqrt n$ nonzero eigenvalues} this sketching matrix preserves the success probability $1/2 + \e$ when $\e \gg 1/\sqrt n$.
To give a tight analysis of quantities like the variance of the outcomes of the protocol after sketching -- i.e., second-moment quantities like $\iprod{(S \tilde{a}_x)(S \tilde{a}_x)^\top, S (A_i \tensor I) S^\top}^2$ -- we use a combinatorial moment-method argument, which crucially uses our bounds on $\|A_i\|_F$.
We also employ a number of tools from random matrix theory -- decoupling inequalities, net-based arguments, and the Hansen-Wright inequality.
For details, see Section~\ref{sec:sketching}.

At the end, we arrive at a $\log n$-qubit pure-state protocol with advantage $\gg 1/\sqrt{n}$, which we have already showed is impossible.
This completes the proof sketch of Lemma~\ref{lem:quantum-lb}, which in turn completes our proof sketch of the moderate-rank Matrix Spencer theorem.

\subsection{Related Work}

\paragraph{Discrepancy}
Discrepancy theory is rich and well explored area of combinatorics with connections to many areas of mathematics and theoretical computer science. 
It has found applications in diverse areas such as approximation algorithms, differential privacy and probability theory. 
For a more thorough introduction, see \cite{chazelle_2000,matousek2009geometric}. 

Classical results in combinatorial discrepancy are often based on linear dependencies (see \cite{barany2008power}) and counting arguments (e.g. Beck's partial coloring method and Spencer's entropy method).
Spencer's six standard deviations theorem was initially proved by combining partial coloring and counting arguments \cite{spencer1985six}. 
Another proof of this theorem was given by \cite{gluskin1989extremal,giannopoulos1997some} by making connections between existence of partial colorings and convex geometry. 
This theorem gave natural conditions to find partial colorings in general convex bodies. 

Successful as they were, these techniques were all non-constructive and thus did not provide algorithmic insights on constructing colorings. 
In a breakthrough result, \cite{bansalspencer} gave the first algorithm to find the signs promised by Spencer's theorem.
The algorithm was based on semidefinite programming but needed to assume the existence of a good coloring in the analysis.
\cite{lovettmeka} gave a random walk-based algorithm whose analysis does not appeal to Spencer's theorem.  
\cite{rothvossconvex} gave an elegant algorithm that produces the partial colorings in convex sets guaranteed by Gluskin's theorem.
\cite{eldan2018efficient} provide an alternative algorithm for this problem using linear programming by providing connections to the width of the convex set -- our SDP-based algorithm is a direct descendent of theirs.

Another line of work in algorithmic discrepancy is constructing algorithms for the \emph{Beck-Fiala} and \emph{Koml\'os} settings, where additional assumptions on the vectors $v_1,\ldots,v_n$ lead to tighter discrepancy bounds.
Here obtaining tight bounds remains an open problem, even non-algorithmically.
The best known non-constructive bounds are obtained using a technique introduced by \cite{banaszczyk1998balancing} which also draws from connections to convex geometry. 
A recent line of work resolved the question of algorithmically matching Banaszczyk's bound \cite{bansal2019algorithm,bansal2018gram, dadush2019towards}. 
For an overview of this line of work, see \cite{garg_2018}. 
    
\paragraph{Matrix discrepancy and spectral graph theory}
Another line of work that is closely related to discrepancy is the construction of sparsifiers for graphs. 
\cite{batson2012twice} construct linear sized (weighted) sparsifiers for graphs that approximate the Laplacian of the graph. 
In a celebrated work, \cite{marcus2015interlacing} use a novel technique based on interlacing polynomials to resolve the Kadison--Singer conjecture, which can be interpreted as a tight discrepancy bound for signed sums of rank-one matrices $a_1 a_1^\top,\ldots,a_n a_n^\top$ in isotropic position.
This is also related to constructing unweighted sparsifiers for graphs. 
It remains an excellent open problem to find algorithms matching the bounds of \cite{marcus2015interlacing}. 

The Matrix Spencer conjecture is a natural matrix generalization of the Spencer theorem, asking if we can improve upon the matrix Chernoff bound to get a bound similar to the one guaranteed by Spencer's theorem.  
It was popularized in a \hyperref{https://windowsontheory.org/2014/02/07/discrepancy-and-beating-the-union-bound/}{category}{name}{blog post} by Raghu Meka. 
This bound was also conjectured in \cite{zouzias2012matrix}. 
\cite{levy2017deterministic} provide an algorithm gives a discrepancy bound of $O\left( \sqrt{n \log q} \right) $ for matrices that are block diagonal with block size $q$. 
A result by \cite{kyng2020four} uses techniques from \cite{marcus2015interlacing} to resolve the conjecture for rank one matrices. 
\cite{reis2020linear} bring together techniques from the convex geometric approach of \cite{gluskin1989extremal,giannopoulos1997some} for constructing algorithms for graph sparsification problems akin to \cite{batson2012twice} towards potentially using these ideas to resolve the matrix Spencer conjecture. 
    
\paragraph{Communication complexity and quantum random access codes}
Lower bounds in one-way communication complexity are widely used to prove lower bounds in other settings: data structures and streaming algorithms, to name just two.
See \cite{rao2020communication} for a modern introduction to communication complexity.
The index function, in particular, plays a central role in one-way communication, see e.g. \cite{kremer1999randomized}.
It is a folklore result that the one-way constant-error classical communication complexity of the $n$-bit index function is $\Omega(n)$.

Quantum protocols for the index function also go by the name \emph{quantum random access codes}, which have been studied intensively in the physics literature, including experimental demonstrations of quantum protocols whose success probabilities are strictly better than those achievable by classical protocols for small $n$, e.g. \cite{tavakoli2015quantum}.
\cite{ambainis2002dense} show that the one-way constant-error quantum communication complexity of the $n$-bit index function is $\Omega(n)$.
This argument was simplified and refined in \cite{nayak1999optimal}.

\section{Preliminaries}
For vectors $x,y \in \R^d$, let $ \iprod{x,y} $ denote the standard inner product $ \sum_i x_i y_i  $. 
For matrices $A,B$, this inner product also corresponds to $ \iprod{A,B} = \Tr\left( AB \right) $. 
Let $ \norm{ x }_2 = \sqrt{ \sum_i x_i^2 }  $ denote the $ \ell_2 $ norm, $ \norm{x}_{\infty} = \max_i \abs{x_i} $ denote the $\ell_{\infty} $ norm and $ \norm{x}_1 = \sum_i \abs{x_i} $ denote the $ \ell_1$ norm. 

For a matrix $A \in \R^{d \times d} $, denote by $ \Norm{A}_F = \sqrt{ \Tr\left(  A^{\top} A \right) } $ the Frobenius norm, by $\Norm{A} =  \sup_{\norm{x} =  1}  \norm{Ax} $ the operator or spectral norm and by $ \norm{A}_1 = \Tr \left( \sqrt{ A^{\top} A } \right) $, the nuclear or the trace norm. 
For matrices $A,B $, $ A \preceq B $ if $ B - A  $ is a positive semidefinite matrix.

A convex set $ \mathcal{K}$ is said to be centrally symmetric if $x \in \mathcal{K} $ implies $ -x \in \mathcal{K} $.
For any convex set $ \mathcal{K}$, denote by $ p_{\mathcal{K}}  $ the Minkowski functional defined by $  p_{\mathcal{K}} \left( x \right) = \inf \left\{ r >0 : x \in r\mathcal{K} \right\}  $.
We say that a convex set has non-empty interior if there is an $ \epsilon >0 $ such that for all $ x $ such that $ \norm{x}_2 \leq \epsilon $, $ x\in K $. 
A compact, convex set set with non-empty interior is referred to as a convex body. 
If $ \mathcal{K} $ is a symmtric convex body, $ p_{\mathcal{K}} $ corresponds to a norm which we denote by $ \norm{ \cdot }_{\mathcal{K}}  $. 
Furthermore, any norm can be seen as the Minkowski functional of its unit ball. 

For any norm $ \norm{\cdot} $, define the dual norm $ \norm{\cdot}_{*} $ by $ \norm{z}_{*} = \sup \left\{ \iprod{x,z} : \norm{x} \leq 1 \right\} $. 
For any convex body $ \mathcal{K}  $ with $ 0 \in \mathcal{K} $, define the polar as $ K^{*} = \left\{ y: \sup_{x \in \mathcal{K}} \iprod{x , y } \leq 1  \right\} $. 
For any symmetric convex body, the dual norm of $ \norm{ \cdot }_{ \mathcal{K} } $ is given by $ \norm{ \cdot }_{\mathcal{K}^{*}  } $.
The dual norm of $ \norm{\cdot}_2 $ is itself, while the dual norm of $ \norm{\cdot}_{\infty} $ is $ \norm{ \cdot}_1 $ (and vice versa). 
For matrix norms, the dual of $ \norm{\cdot}_{F} $ is itself, while the dual norm of $ \norm{\cdot} $ is $ \norm{\cdot}_1 $ (and vice versa). 

For any random variable $X$, let $ \E X $ denote its expectation (if it exists) and let $ \Var\left( X \right) = \E X^2 - \left( \E X \right)^2 $ denote the variance (if it exists).
For any $ \mu \in \R^d $ and $ \Sigma \succeq 0  $, denote by $ \cN\left( \mu , \Sigma \right) $, the normal distribution with mean $ \mu  $ and covariance $ \Sigma $.

We also record here some notation from quantum information. 
A density matrix $ \rho $ is a positive semidefinite matrix with trace one i.e. $ \rho \succeq 0 $ and $ \tr \rho = 1 $. 
Measurements in quantum information are specified by POVMs which are PSD matrices $A_i$ such that $ \sum_i A_i = I $. 
For any density matrix, upon measuring $ \rho $ with respect to the POVM $ \left\{ A_i \right\} $, one gets outcome $i$ with probability $ \tr \rho A_i $. 
For any density matrix, define the von Neumann entropy as $ S\left( \rho \right) = \tr \left( \rho \log \rho \right) $.

\section{Proof of Main Theorem}

In this section we prove the following main partial coloring theorem.
Then in Section~\ref{sec:matrix-spencer-moderate} we use it to deduce Theorem~\ref{thm:matrix-spencer-intro}.
Theorem~\ref{thm:full-spencer-intro} can then be proved by a simple modification of the proof of Theorem~\ref{thm:matrix-spencer-intro}.

\begin{theorem}[Main Partial Coloring Theorem]
    \label{thm:main-partial-coloring}
    Let $A_1,\ldots,A_n \in \R^{d \times d}$ be symmetric.
    There is a partial fractional coloring $x \in [-1,1]^n$ such that $\Omega(n)$ indices $i$ have $|x_i| = 1$ and
    \[
        \Norm{\sum_{i \leq n} x_i A_i} \leq \Norm{ \sum_{i \leq n} A_i^2 }^{1/2} \cdot
        O\Paren{\sqrt{ 1 + \log \frac{\Tr \sum_{i \leq n} A_i^2}{\sqrt{n} \Norm{\sum_{i \leq n} A_i^2}} }} \mper
    \]
    Furthermore, there is a randomized polynomial time algorithm which finds such a coloring with high probability.
\end{theorem}

It is a folklore observation\footnote{Thanks to Raghu Meka for making us aware of this.} that if $n \gg d^2$ one can find a partial coloring with zero discrepancy by linear programming, so Theorem~\ref{thm:main-partial-coloring} is interesting when $n \ll d^2$.

We now assemble our main tools for the proof of Theorem~\ref{thm:main-partial-coloring}.
Our first lemma shows that if a low-discrepancy partial fractional coloring of $A_1,\ldots,A_n$ does not exist then $A_1,\ldots,A_n$ induce a scheme to compress a large subset of $\{ \pm 1\}^n$ into the $d \times d$ nuclear norm ball.
The nuclear norm appears because it is dual to the norm in which we are measuring discrepancy, namely spectral norm.
Using convex duality, in Section~\ref{sec:compress-or-color} we actually prove the following more general statement which applies to any norm and its dual, in hope that it is useful in future work.

\begin{lemma}[Compress or Color]
    \label{lem:compress-or-color}
  Let $\cK$ be a symmetric convex body in $\R^m$ and let $\|\cdot\|_{\cK}$ be its associated norm.
    Let $v_1,\ldots,v_n \in \R^m$.
    For every $\e, \Delta > 0$, either
  \begin{itemize}
      \item there is a partial fractional coloring $x \in [-1,1]^n$ such that $| \{ i \, : \, |x_i| = 1 \}| \geq \e n$ and $\| \sum_{i \leq n} x_i v_i \|_{\cK} \leq \Delta$, and, furthermore, with probability $\Omega(1/n)$ over uniformly random choice of $g \in \{ \pm 1\}^n$ such a coloring is given by optimizer of the following convex program:
        \[
            \max_x \iprod{x,g} \text{ such that } \Norm{\sum_{i \leq n} x_i v_i}_{\cK} \leq \Delta \text{ and } x \in [-1,1]^n \mcom \text{ or,}
        \]
    \item for at least $2^n/2$ choices of $g \in \{\pm 1\}^n$ there is a vector $y_g \in \R^m$ with $\|y_g\|_{\cK^{*}} = 1$, a set $I_g \subseteq [n]$ with $|I_g| \leq \e n$, and for $i \in [n] \setminus I_g$ numbers $\Delta_{ig} \geq \Omega(\Delta)$ such that $\iprod{y_g, v_i} = \tfrac{\Delta_{ig}}{n}g_i$.
  \end{itemize}
\end{lemma}

In light of Lemma~\ref{lem:compress-or-color}, to show in the proof of Theorem~\ref{thm:main-partial-coloring} that there is a partial fractional coloring with discrepancy $\Delta$, we can instead rule out a mapping from $g \in \{ \pm 1\}^n$ to matrices $Y_g$ with $\|Y_g\|_1 = 1$ (since the nuclear norm is dual to the spectral norm) such that $\iprod{Y_g,A_i} \approx \tfrac \Delta n \cdot g_i$ for at least $(1-\e)n$ indices $i \in [n]$.

By a simple transformation of the $Y_g$s and $A_i$s, we can assume $Y_g \succeq 0$, and hence that $Y_g$ is a $d$-dimensional density matrix, and that $\Tr A_i = 0$.
We can then interpret $\{Y_g\}_{g \in \{ \pm 1\}^n}$ as a strategy for Alice in a one-way quantum protocol for the $n$-bit index function, where Bob's measurements are the $A_i$'s.
In our discussion of such protocols so far, we have always assumed that $\|A_i\| \leq 1$; note that we do not make this assumption here.
It turns out that the weaker assumption on $\sum_{i \leq n} A_i^2$ suffices to build the repeated protocol we need to prove our communication lower bound.

Since our communication lower bounds only apply to protocols where Alice communicates a pure state, we use following lemma to round $\{Y_g\}_{g \in \{ \pm 1\}^n}$ to pure states $\{y_g\}_{g \in \{ \pm 1 \}^n}$.
The cost is that the fluctuations in this randomized rounding scheme are governed by $\Tr \sum_{i \leq n} A_i^2 = \sum_{i \leq n} \|A_i\|_F^2$ in addition to $\| \sum_{i \leq n} A_i^2 \|$.
The assumption $\|A_i\|_F \leq n^{1/4}$ is needed to control the first term.

\begin{lemma}[Purify then sketch]
  \label{lem:purify-then-sketch}
    Let $n,d,\delta > 0$.
    Let $A_1,\ldots,A_n \in \R^{d \times d}$ have $\Tr A_i = 0$ and $A = \sum_{i =1}^n A_i^2$.
    Let $\{Y_g\}_{g \in \{ \pm 1\}^n }$ be density matrices.

    For every integer $r > 0$, there exist symmetric matrices $B_1,\ldots,B_n \in \R^{r \times r}$ such that 
    \[
        \Norm{\sum_{i=1}^n B_i^2} \leq O_\delta \Paren{\|A\| + \frac{\min(\sqrt{n},d) \cdot \Tr A}{r}}
    \]
    and such that for at least $\tfrac 3 4 \cdot 2^{n}$ of $g \in \{ \pm 1\}^n$ there exists an $r$-dimensional pure state $y_g$ (i.e. a vector $y_g \in \R^r$ with $\|y_g\| = 1$) and a number $c_g \geq \Omega_\delta(1)$ such that for at least $(1-\delta)n$ indices $i \in [n]$,
    \[
        \Abs{\iprod{y_g y_g^\top, B_i} - c_g \iprod{Y_g,A_i} } \leq O_\delta\Paren{\frac{\|A\|}{nr} + \frac{ \min(\sqrt{n},d) \Tr A}{n r^2}}^{1/2} \mper
    \]
\end{lemma}

Lastly, we prove the following lemma using tools from quantum information and communication complexity -- see Section~\ref{sec:qrac-lb} for a more thorough discussion.

\begin{lemma}
    \label{lem:rac-lb-matrix-dependent}
    There is a universal constant $\delta > 0$ such that the following holds for all integers $n,m > 0$.
    Let $\cS \subseteq \{ \pm 1\}^n$ have size at least $2^{\delta n}$.
    Suppose that $\{y_g\}_{g \in \cS}$ is a collection of $m$-qubit pure states such that for some symmetric matrices $A_1,\ldots,A_n$, subsets $I_g \in \binom{n}{\delta n}$ and numbers $\{\eta_{ig} > 0\}_{g \in \cS, i \in [n] \setminus I_g}$, for all $i \in [n] \setminus I_g$ it holds that $\iprod{y_gy_g^\top, A_i} = \eta_{ig} \cdot g_i$.
    Let 
    \[
        \eta^2 = \frac{\E_{g \sim \cS, i \sim [n] \setminus I_g} \eta_{ig}^2 } {\|\E_{i \sim [n]} A_i^2 \|}\mper
    \]
    If $\eta \geq C / \sqrt{n}$ for some universal $C > 0$, then
    \[
        m \geq \log \frac 1 {\eta^2} + \Omega( \eta^2 \cdot n)\mper
    \]
\end{lemma}

With our tools in hand, we can prove Theorem~\ref{thm:main-partial-coloring}.

\begin{proof}[Proof of Theorem~\ref{thm:main-partial-coloring}]
    Let
    \begin{quote}
        \textbf{Good Compression$(\Delta)$:} for $2^n/2$ choices of $g \in \{ \pm 1\}^n$ there is a matrix $Y_g$ with $\|Y_g\|_1 = 1$, a set $I_g \subseteq [n]$ with $|I_g| \leq \e n$, and numbers $\{\Delta_{ig}\}_{i \in [n] \setminus I_g}$ with $\Delta_{ig} \geq \Delta$ such that $\iprod{Y_g, A_i} = \tfrac{\Delta_{ig}}{n} g_i$ for $i \in [n] \setminus I_g$
    \end{quote}

    By the Compress or Color Lemma~\ref{lem:compress-or-color}, it is enough to show that for a small-enough constant $\e > 0$, if \textbf{Good Compression$(\Delta)$} occurs, then
    \begin{align}
        \label{eq:main-1}
        \Delta \leq \Norm{ \sum_{i \leq n} A_i^2 }^{1/2} \cdot
        O\Paren{\min \left \{ \sqrt{ 1 + \log \frac{\Tr \sum_{i \leq n} A_i^2}{\sqrt{n} \Norm{\sum_{i \leq n} A_i^2}} }, \sqrt{n \cdot 2^{-\Omega(n/d^2)}} \right \}} \mper
    \end{align}
    In that case, if $\Delta$ is larger than in \eqref{eq:main-1}, then with probability $\Omega(1/n)$ over choice of $g \sim \{ \pm 1\}^n$, the semidefinite program
    \[
        \max_{x \in [-1,1]^n} \iprod{x,g} \text{ such that } \Norm{\sum_{i \leq n} x_i A_i} \leq \Delta
    \]
    finds a fractional partial coloring with $\e n$ integer entries and discrepancy at most $\Delta$.

    Suppose \textbf{Good Compression$(\Delta)$} occurs, for some $\Delta > 0$.
    Let $Y_g^+ \succeq 0$ be the positive semidefinite part of $Y_g$ and $Y_g^- \preceq 0$ the negative definite part, so that $Y_g = Y_g^+ + Y_g^-$.
    By replacing $Y_g$ and $A_i$ with the following block matrices:
    \[
        Y_g \rightarrow \Paren{\begin{array}{cc}Y_g^+ & 0 \\ 0 & -Y_g^- \end{array} } \text{ and } A_i \rightarrow \Paren{ \begin{array}{cc} A_i & 0 \\ 0 & -A_i \end{array} }\mcom
    \]
    (and replacing $d$ with $2d$) we may assume that $Y_g \succeq 0$ with $\Tr Y_g = 1$ and $\Tr A_i = 0$.
    Note that $\iprod{Y_g, A_i}$ and $\|\sum_{i \leq n} A_i^2\|$ are preserved by this transformation, and $\Tr \sum_{i \leq n} A_i^2$ grows by a factor of $2$.

    Let $A = \sum_{i \leq n} A_i^2$. By the Purify-then-Sketch lemma~\ref{lem:purify-then-sketch}, for any choice of integer $r > 0$ and any $\delta > 0$ there are symmetric matrices $B_1,\ldots,B_n \in \R^{r \times r}$ such that 
    \begin{align*}
        \Norm{\sum_{i \leq n} B_i^2 } \leq O_\delta \Paren{ \|A\| + \frac{\min(\sqrt{n},d) \Tr A}{r}}
    \end{align*}
    and for at least $\tfrac 3 4 \cdot 2^n$ choices of $g \in \{ \pm 1\}^n$ there is a pure state $y_g \in \R^r$ and a number $c_g \geq \Omega_\delta(1)$ such that for at least $(1-\delta)n$ indices $i \in [n]$,
    \[
        \Abs{ \iprod{y_gy_g^\top, B_i} - c_g \iprod{Y_g, A_i} } \leq O_\delta \Paren{ \frac{ \|A\|}{nr} + \frac{\min(\sqrt{n},d) \Tr A}{nr^2}}^{1/2}\mper
    \]
    Now, if 
    \begin{align}
        \label{eq:main-2}
        \Delta^2 \geq O_\delta(1) \cdot \|A\| \cdot \max\Paren{\frac n r, \frac n r \cdot \frac{\min(\sqrt n, d)\Tr A}{r}}\mcom
    \end{align}
    then for at least $2^n/4$ choices of $g$ there is a subset $I_g \subseteq [n]$ with $|I_g| \leq \e + \delta$ such that for $i \in [n] \setminus I_g$,
    \[
        \iprod{y_g y_g^\top, B_i} = \frac{\Delta_{ig}}{n} \cdot g_i
    \]
    where $\Delta_{ig} \geq \Omega(\Delta)$.
    Let us call this set of $g$'s $\cS$.

    Let
    \[
        \eta^2 = \frac{\E_{g \sim \cS, i \sim [n] \setminus I_g} (\Delta_{ig}/n)^2}{\Norm{\E_{i \sim [n]} B_i^2}}\mper
    \]
    Henceforth taking $\delta$ to be a small-enough universal constant, by Lemma~\ref{lem:rac-lb-matrix-dependent}, either $\eta \leq O(1/\sqrt{n})$ or $r \geq 2^{\Omega(\eta^2 n)} \cdot \tfrac 1 {\eta^2}$.
    We treat the two cases separately.

    \subparagraph{Case 1A: $\eta \leq O(1/\sqrt{n})$}
    Using $\Delta_{ig} \geq \Omega(\Delta)$, in this case we have
    \[
        \Delta^2 \leq O(n) \cdot \Norm{\E_{i \sim [n]} B_i^2} \leq O \Paren{ \|A\| + \frac{\min(\sqrt{n},d) \Tr A}{r}}
    \]

    \subparagraph{Case 1B: $r \geq 2^{\Omega(\eta^2 n)} \cdot \tfrac 1 {\eta^2}$}
    Using the definition of $\eta$, we have
    \[
        \log \Paren{ r \cdot \frac{\E_{g \sim \cS, i \sim [n] \setminus I_g} (\Delta_{ig}/n)^2}{\Norm{\E_{i \sim [n]} B_i^2}}} \geq \Omega(n) \cdot \frac{\E_{g \sim \cS, i \sim [n] \setminus I_g} (\Delta_{ig}/n)^2}{\Norm{\E_{i \sim [n]} B_i^2}}
    \]
    This rearranges to
    \[
        \log \Paren{ \frac r n \cdot \frac{\E_{g \sim \cS, i \sim [n] \setminus I_g} \Delta_{ig}^2}{\Norm{\sum_{i \leq n} B_i^2}}} \geq \Omega(1) \cdot \frac{\E_{g \sim \cS, i \sim [n] \setminus I_g} \Delta_{ig}^2}{\Norm{\sum_{i \leq n} B_i^2}}
    \]
    which gives
    \[
        \log \frac r n \geq \Omega(1) \cdot \frac{\E_{g \sim \cS, i \sim [n] \setminus I_g} \Delta_{ig}^2}{\Norm{\sum_{i \leq n} B_i^2}}\mcom
    \]
    so, rearranging and using our bound on $\|\sum_{i \leq n} B_i^2\|$, and that $\Delta_{ig} \geq \Omega(\Delta)$, we get
    \[
        \Delta^2 \leq O\Paren{ \|A\| + \frac{\min(\sqrt n, d) \Tr A}{r} } \cdot \log \frac r n\mper
    \]
   
    Now let us choose 
    \[
        r = \max \Paren{n, \frac{\min(\sqrt{n}, d) \Tr A}{\|A\|}}\mcom
    \]
    so that putting together \eqref{eq:main-2} with cases 1A and 1B, we find
    \[
        \Delta^2 \leq O(\|A\|) \max \Paren{1, \log \frac{\min(\sqrt n, d) \Tr A}{n \|A\|}}\mper 
    \]

%
\end{proof}

\subsection{Matrix Spencer for Moderate-Rank Matrices}
\label{sec:matrix-spencer-moderate}

In this section we use Theorem~\ref{thm:main-partial-coloring} to prove the following theorem.

\begin{theorem}
    Let $ d \geq n $ and $A_1,\ldots,A_n \in \R^{d \times d}$ be symmetric matrices with $\|A_i\| \leq 1$ and $\|A_i\|_F^2 \leq \sqrt{d}$ for all $i \in [n]$.
      There is a coloring $x \in \{ \pm 1\}^n$ such that
      \[
          \Norm{\sum_{i \leq n} x_i A_i} \leq O\left(\sqrt{n \log \left( \frac{d}{n} \right)  }\right)\mper
      \]
      Furthermore, there is a (randomized) polynomial time algorithm which finds such a coloring with high probability.
  \end{theorem}

\begin{proof}
    We will get a full coloring for this setting by iteratively applying Lemma~\ref{thm:main-partial-coloring}. 
    Consider the first round of partial coloring.
    Note that since $ \Norm{A_i} \leq 1 $, we have $ \Norm{ \sum_{i\leq n} A_i^2  } \leq n  $.
    From Lemma~\ref{thm:main-partial-coloring}, we get that there is a partial coloring $x$ with $ c n $ co-ordinates such that $ x_i \in \left\{ -1,1 \right\}$ and 
    \begin{equation*}
        \Norm{\sum_{i \leq n} x_i A_i} \leq \Norm{ \sum_{i \leq n} A_i^2 }^{1/2} \cdot
        O\Paren{ \sqrt{ 1 + \log \frac{\Tr \sum_{i \leq n} A_i^2}{\sqrt{n} \Norm{\sum_{i \leq n} A_i^2} }  } } \mper
    \end{equation*}
    By hypothesis on $\|A_i\|_F^2 = \Tr A_i^2$, we have $\Tr \sum_{i \leq n} A_i^2 \leq n\cdot d^{1/2}$. 
    \[
        \Norm{\sum_{i \leq n} x_i A_i} \leq O\Paren{\Norm{\sum_{i \leq n} A_i^2}}^{1/2} \cdot \sqrt{ \log \frac{ 2 \sqrt{dn}  }{\Norm{\sum_{i \leq n} A_i^2}} } \mper
    \]

    Since $x \log ( 2\sqrt{dn} /x) \leq O \left( n \log( d/n )  \right) $ for $0 \leq x \leq n$, 
    we get a partial coloring with discrepancy $ O \left( \sqrt{n \log(d/n) } \right)  $. 

    Given a partial coloring $x$, we move to the next round of partial coloring by replacing $A_i$ by 
    \begin{equation*}
        A'_i =  \sign \left( x_i \right) \cdot  \left( 1 - \abs{x_i} \right) A_i. 
    \end{equation*}
    We ignore the co-ordinates corresponding to zero matrices. 
    Since $ \left( 1 - \abs{x_i} \right) \leq 1 $, the new matrices still satisfy the requirements on the spectral norm and Frobenius norm.  
    Furthermore, since $ cn $ co-ordinates we integral, the number of matrices is now $ \left( 1 - c \right) n $. 
    Thus, we get a partial coloring with $ y $ such that 
    \begin{equation*}
        \Norm{ \sum y_i  A'_i } \leq \Norm{ \sum_{i} A_i^{'2} }^{1/2}  O\Paren{ 1 + \log \frac{\Tr \sum_{i } A_i^{'2}}{\sqrt{\left( 1 - c \right)n} \Norm{\sum_{i} A_i^{'2} } }  }
    \end{equation*} 
    Arguing as before, noting that $ \Tr \sum_i A_i^{'2} \leq \left( 1 - c \right)n \sqrt{d}   $, we get 
    \begin{equation*}
        \Norm{ \sum y_i  A'_i } \leq O \left( \sqrt{ \left( 1 - c \right) n  \log\left( \frac{ d}{ (1-c) n  } \right) }    \right)   .
    \end{equation*}
    
    Then, consider the partial coloring with co-ordinates $ z_i = x_i +  \sign(x_i) \left( 1 - \abs{x_i} \right) y_i    $. 
    First note that we have $ c + \left( 1-c \right)c/2 $ co-ordinates that are integral. 
    To see this note that for half the integral co-ordinates in $y$, we must have $ \sign\left( y_i \right) =1 $ (replacing $y$ with $ -y$ if necessary). 
    For such $y_i$, we have $ x_i + \sign\left( x_i \right) \left( 1 - \abs{x_i} \right) y_i = \sign\left( x_i \right)  $ which is integral. 
    Furthermore, 
    \begin{equation*}
        \Norm{ \sum_i z_i A_i  } \leq \Norm{  \sum_i x_i A_i  } + \Norm{  \sum_i  y_i A'_i } \leq  O\left(  \sqrt{n \log \left( \frac{d}{n} \right) }  + \sqrt{ \left( 1-c \right) n   \log\left( \frac{d}{  (1-c) n  } \right) } \right).  
    \end{equation*} 
    Iterating this inductively, we get that the discrepancy is bounded by 
    \begin{equation*}
        O\left(  \sqrt{n} \sum_{i}  \left( 1 - c \right)^{i/2} \sqrt{ \log\left( \frac{d}{ \left( 1 - c \right)^i n } \right) }  \right) = O\left( \sqrt{n \log \left( \frac{d}{n} \right) }  \right) 
    \end{equation*}
    as required.


    
\end{proof}

\section{Compress or Color}
\label{sec:compress-or-color}


In this section we prove Lemma~\ref{lem:compress-or-color}.
The lemma follows from the following two propositions connecting the ``compress'' and ``color'' cases both to the Rademacher width of the set of partial fractional colorings.
We use the following notation: for a set $K \subseteq \R^n$ and $I \subseteq [n]$, let
\[
    K\left( I \right) = K \cap \{x \, : \, x_i \in [-1,1] \text{ for } i \in I \}  \mper
\]

    First, we give a sufficient condition for having a partial coloring with a large fraction of its coordinates integral. 
    The proof follows ideas from \cite{eldan2018efficient}. 
    \begin{proposition}
        \label{prop:width-to-coloring}
        Let $ K \subseteq \R^n $ be a convex set such that 
            \begin{equation*}
                \E_{g} \min_{\substack{I \subseteq [n] \\ |I| = \e n}} \max_{ x \in K \left( I \right) } \iprod{ x , g}  = cn 
            \end{equation*}
            for some $c>1$.
            Let $R = \max_{x \in K} \|x\|$ be the diameter of $K$.
            If $y = y(g) = \argmax_{x \in K([n]) } \iprod{x , g} $, then with probability at least $(c-1)^2 n/ R^2$ over uniformly random $g \sim \{ \pm 1\}^n$ we have $ \abs{\left\{ i : \abs{y_i} = 1 \right\} } \geq \epsilon n$. 
    \end{proposition}
    \begin{proof}
    
        First note that $  \max_{x \in  K([n]) }  \iprod{x , g} \leq \max_{x \in  [-1,1]^n }  \iprod{x , g } \leq n  $. 
        For a given $g \in \left\{ -1,1 \right\}^n $, let $I_g$ denote the set of coordinates of $ y = \argmax_{x \in K([n]) } \iprod{x , g } $ that are in $ \left\{ -1,1 \right\} $. 
        We would like to bound the probability that $ \abs{I_g} \geq \epsilon n  $. 
        Consider 
        \begin{align*}
            \E_{g} \min_{\substack{I \subseteq [n] \\ |I| = \e n}} \max_{ x \in K \left( I \right) } \iprod{ x , g} = \E_{g} \left[ \mathbb{I} \left[ \abs{I_g} \geq \epsilon n  \right]  \min_{\substack{I \subseteq [n] \\ |I| = \e n}} \max_{ x \in K \left( I \right) } \iprod{ x , g} \right] + \E_{g} \left[ \mathbb{I} \left[ \abs{I_g} < \epsilon n  \right]  \min_{\substack{I \subseteq [n] \\ |I| = \e n}} \max_{ x \in K \left( I \right) } \iprod{ x , g} \right]
        \end{align*}
        Looking at the second term, we get
        \begin{align*}
            \E_{g} \left[ \mathbb{I} \left[ \abs{I_g} <  \epsilon n  \right]  \min_{\substack{I \subseteq [n] \\ |I| = \e n}} \max_{ x \in K \left( I \right) } \iprod{ x , g} \right] & \leq \E_{g} \left[ \mathbb{I} \left[ \abs{I_g} < \epsilon n  \right] \max_{ x \in K \left( I_g \right) } \iprod{ x , g} \right] \\ 
            & = \E_{g} \left[ \mathbb{I} \left[ \abs{I_g} < \epsilon n  \right] \max_{ x \in K([n]) } \iprod{ x , g} \right] \\ 
            & \leq \sqrt{  \E_{g} \left[ \mathbb{I} \left[ \abs{I_g} < \epsilon n  \right] ^2  \right] } \cdot \sqrt{  \E_{g} \left[ \max_{ x \in K([n])  } \iprod{ x , g} ^2  \right] } \\ 
            & \leq n . 
        \end{align*}
        For the equality above, we have used that dropping the constraints $x_i \in [-1,1]$ for $i$ such that the optimizer $x$ of $\max_{x \in K([n])} \iprod{x,g}$ has $-1 < x_i < 1$ does not affect the maximum value.
        Looking at the first term, we have 
        \begin{align*}
            \E_{g} \left[ \mathbb{I} \left[ \abs{I_g} \geq \epsilon n  \right]  \min_{\substack{I \subseteq [n] \\ |I| = \e n}} \max_{ x \in K \left( I \right) } \iprod{ x , g} \right] &\leq \sqrt{ \E_{g} \left[ \mathbb{I} \left[ \abs{I_g} \geq \epsilon n  \right]^2 \right]   } \cdot \sqrt{   \E\left[  \min_{\substack{I \subseteq [n] \\ |I| = \e n}} \max_{ x \in K \left( I \right) } \iprod{ x , g}^2  \right] } \mper
        \end{align*}
        Putting these together and using $c > 1$, we get
        \[
 \Pr_g \Paren{\abs{I_g} \geq \e n} \geq \frac{(c-1)^2n^2}{\E\left[  \min_{\substack{I \subseteq [n] \\ |I| = \e n}} \max_{ x \in K \left( I \right) } \iprod{ x , g}^2  \right] }\mper
        \]
        To bound the denominator, we observe that for all $x \in K$ and all $g \in \{ \pm 1\}^n$ we have $\iprod{x,g}^2 \leq \|x\|^2 \|g\|^2 \leq R^2 n$.
        All in all, we have
        \[
            \Pr_g \Paren{\abs{I_g} \geq \e n} \geq \frac{(c-1)^2 n}{R^2}\mper \qedhere
        \]
    \end{proof}

    Using the contrapositive of Proposition~\ref{prop:width-to-coloring} in the context of Lemma~\ref{lem:compress-or-color}, with $R = \Theta(n)$, if the ``furthermore'' portion of the first condition fails, then $ \E_{g} \min_{\substack{I \subseteq [n] \\ |I| = \e n}} \max_{ x \in K \left( I \right) } \iprod{ x , g}  \leq 2n $, where $K = \{ x \, : \, \norm{ \sum x_i v_i} \leq \Delta \}$. 
    In the next Proposition, we use convex programming duality to show that this implies the second condition in Lemma~\ref{lem:compress-or-color}.

\begin{proposition}
    \label{prop:width-to-compression}
    Let $\cK,\|\cdot\|_{\cK},$ and $v_1,\ldots,v_n$ be as in Lemma~\ref{lem:compress-or-color}.
    For $\Delta > 0$, let
    \[
        K_\Delta = \left \{x \in \R^n \, : \, \Norm{\sum_{i \leq n} x_i v_i }_{\cK} \leq \Delta, \|x\|_2 \leq R \right \}
    \]
    be the set of partial fractional colorings of $v_1,\ldots,v_n$ with discrepancy at most $\Delta$ and bounded $\ell_2$ norm.
    Suppose for some $\e > 0$ that
    \[
        \E_{g \sim \{ \pm 1\}^n} \min_{\substack{I \subseteq [n] \\ |I| = \e n}} \max_{x \in K_\Delta(I)} \iprod{x,g} = c n \leq \frac R 4\mper
    \]
    Then for at least $2^n/2$ choices of $g \in \{ \pm 1 \}^n$ there exists $y_g \in \R^m$ with $\|y_g\|_{\cK^*} = 1$, a set $I_g \subseteq [n], |I| = \epsilon n$, and numbers $\Delta_{ig} \geq \Delta / (8c)$ for $i \in [n] \setminus I_g$  such that $\iprod{y_g,v_i} = \tfrac {\Delta_{ig}}{n}g_i$.
\end{proposition}
\begin{proof}
    Fix $g \in \{ \pm 1\}^n$.
    Suppose $\min_{I \subseteq [n], |I| = \e n} \max_{x \in K_\Delta(I)} \iprod{x,g} = c_g n$ for some $c_g \geq 0$.
    Let $I_g \subseteq [n], |I_g| = \e n$ witness this.
    Consider the convex program 
    \begin{equation*}
        \max_x \iprod{x,g} \text{ such that } \Norm{\sum_{i \leq n} x_i v_i}_{\cK} \leq \Delta, \|x\|_2 \leq R, \text{ and } x_i \in [-1,1] \text{ for all } i \in I_g\mper
    \end{equation*}
    We first note that strong duality holds for this convex program by using Slater's condition and noting that $0$ is a strictly feasible point as $\cK$ has non-empty interior. 
    Writing the Lagrangian with dual variables $\alpha, \{\beta_i^-,\beta_i^+\}_{i \in I_g}, \gamma$, we get 
    \begin{align*}
        & \max_{x}  \min_{\alpha,\beta^{+},\beta^{-},\gamma \geq 0 }  \iprod{x , g} + \alpha \left( \Delta - \Norm{ \sum_{i\leq n} x_i v_i   }_{\cK} \right) +  \sum_{i \in I_g} \beta^{+}_i \left( 1- x_i \right) + \sum_{i \in I_g}\beta^{-}_i  \left( 1+ x_i \right) + \gamma \Paren{R - \|x\|_2}\\ 
        & = \max_{x}  \min_{\alpha,\beta^{+},\beta^{-},\gamma \geq 0 }  \iprod{x , g} + \alpha \left( \Delta - \max_{ \norm{y}_{\cK^{*} } \leq 1 } \iprod{ \sum_{i\leq n} x_i v_i ,y  } \right) +  \sum_{i \in I_g} \beta^{+}_i \left( 1- x_i \right) + \sum_{i \in I_g}\beta^{-}_i  \left( 1+ x_i \right) + \gamma \Paren{R - \max_{\|z\|_2 \leq 1} \iprod{x,z} } \\ 
        & = \max_{x}  \min_{\alpha,\beta^{+},\beta^{-},\gamma \geq 0 } \min_{ \norm{y}_{\cK^{*}} \leq 1 } \min_{\|z\|_2 \leq 1}  \iprod{x , g} + \alpha \left( \Delta -  \iprod{ \sum_{i\leq n} x_i v_i ,y  } \right) +  \sum_{i \in I_g} \beta^{+}_i \left( 1- x_i \right) + \sum_{i \in I_g}\beta^{-}_i  \left( 1+ x_i \right) + \gamma \Paren{R - \iprod{x,z}}\mper 
    \end{align*}
    Applying Slater's condition, we get that dual
    \begin{align*}
        \min_{\alpha,\beta^{+},\beta^{-},\gamma \geq 0 }   \max_{x}  \min_{ \norm{y}_{\cK^{*}, \|z\|_2} \leq 1 }   \iprod{x , g} + \alpha \left( \Delta -  \iprod{ \sum_{i\leq n} x_i v_i ,y  } \right) +  \sum_{i \in I_g} \beta^{+}_i \left( 1- x_i \right) + \sum_{i \in I_g}\beta^{-}_i  \left( 1+ x_i \right) + \gamma\Paren{R - \iprod{x,z}}
    \end{align*} 
    has the same value as the primal.
    Furthermore, for any $ \alpha,\beta^{+},\beta^{-},\gamma \geq 0 $, consider 
    \begin{equation*}
        \max_{x}  \min_{ \norm{y}_{\cK^{*}} \leq 1, \|z\|_2 \leq 1 }   \iprod{x , g} + \alpha \left( \Delta -  \iprod{ \sum_{i\leq n} x_i v_i ,y  } \right) +  \sum_{i \in I_g} \beta^{+}_i \left( 1- x_i \right) + \sum_{i \in I_g}\beta^{-}_i  \left( 1+ x_i \right) + \gamma \Paren{R - \iprod{x,z}} \mper
    \end{equation*}

    Below, we note general conditions under which we can switch the max and the min in the above expression. 
    \begin{fact}[Sion Minimax Theorem]\label{fact:Sion}
        Let $V_1 $ and $V_2$ be two real topological vector spaces and let $X \subseteq V_1$ and $Y \subseteq V_2$ be convex. 
        Let $ \alpha : X \times Y \to {\mathbb{R}} $ be semicontinuous. 
        Furthermore, for all $x \in X$, let $\alpha \left( x , \cdot \right)$ be quasiconcave and for all $y \in Y$, let $ \alpha\left( \cdot , y \right) $ be quasiconvex.
        Then, if either $X$ or $Y$ is compact, then 
        \begin{equation*}
            \inf_{x \in X} \sup_{y\in Y} \alpha\left( x,y  \right) = \sup_{y\in Y} \inf_{x \in X}  \alpha\left( x,y  \right) . 
        \end{equation*}
    \end{fact}
    Applying Fact~\ref{fact:Sion} while noting that $ \cK^{*}$ and the Euclidean ball are compact, we get  
    \begin{align*}
        \min_{\alpha,\beta^{+},\beta^{-},\gamma \geq 0 } \min_{ \norm{y}_{\cK^{*}}, \|z\|_2 \leq 1 }  \max_{x}     \iprod{x , g} + \alpha \left( \Delta -  \iprod{ \sum_{i\leq n} x_i v_i ,y  } \right) +  \sum_{i \in I_g} \beta^{+}_i \left( 1- x_i \right) + \sum_{i \in I_g}\beta^{-}_i  \left( 1+ x_i \right) + \gamma\Paren{R - \iprod{x,z}} \\
        = \min_{\alpha,\beta^{+},\beta^{-},\gamma \geq 0 }   \max_{x}  \min_{ \norm{y}_{\cK^{*}}, \|z\|_2 \leq 1 }   \iprod{x , g} + \alpha \left( \Delta -  \iprod{ \sum_{i\leq n} x_i v_i ,y  } \right) +  \sum_{i \in I_g} \beta^{+}_i \left( 1- x_i \right) + \sum_{i \in I_g}\beta^{-}_i  \left( 1+ x_i \right) + \gamma \Paren{R - \iprod{x,z}}.
    \end{align*}
    So there exist a $ \alpha , \beta^{+} , \beta^{-},\gamma \geq 0 $ and $y $ with $ \norm{y}_{ \cK^{*} } \leq 1  $ and $z$ with $\|z\|_2 \leq 1$, such that for all $x$,  
    \begin{equation} \label{eq:minmax}
        \iprod{x , g} + \alpha \left( \Delta -  \iprod{ \sum_{i\leq n} x_i v_i ,y  } \right) +  \sum_{i \in I_g} \beta^{+}_i \left( 1- x_i \right) + \sum_{i \in I_g}\beta^{-}_i  \left( 1+ x_i \right) + \gamma \Paren{R - \iprod{x,z}} \leq c_g n .
    \end{equation}
    We first claim that for all $x$ such that $x_i = 0$ for all $i \notin I_g$ we have 
    \begin{equation*}
        \iprod{x,g} - \alpha \iprod{y, \sum x_i v_i } - \gamma \iprod{x,z} = 0.
    \end{equation*}
    If this fails for some $x$, then by scaling the maximum would be unbounded, contradicting \ref{eq:minmax}. 
    Setting $ x = e_i $ for each $i \notin I_g $ and plugging in to the above expression, we get 
    \begin{equation*}
        \iprod{y , v_i} = \frac{1}{ \alpha }(g_i - \gamma z_i) \mper
    \end{equation*}
    Furthermore, setting $x = 0$, we get 
    \begin{equation*}
        \alpha \Delta + \sum_{i \in I_g} \beta_i^{+} + \sum_{i \in I_g} \beta_i^{-} + R \gamma \leq c_g n  
    \end{equation*}
    and thus both $\alpha \Delta \leq c_g n$ and $\gamma R \leq c_g n$. 
    So,
    \[
        \iprod{ \frac y {\|y\|_{\cK^*}}, v_i} = \frac 1 { \alpha \|y\|_{\cK^*} } \cdot (g_i -\gamma z_i) = \frac {\Delta_g} n \cdot (g_i - \gamma z_i)
    \]
    where we have set $\Delta_g = n / \alpha \|y\|_{\cK^*}$.
    Since $\|y\|_{\cK^*} \leq 1$, we have $\Delta_g \geq n/\alpha \geq \Delta / c_g$.
    By Markov's inequality, $\Pr(c_g \leq 2 \E_g c_g) \geq 1/2$.
    For each such $g$, we have $\Delta_g \geq \Delta / (2 \E_g c_g)$.

    Finally, $\gamma \leq c_g n / R$ and $|z_i| \leq 1$, so for each $g$ and $i \notin I_g$ we have $|\gamma z_i| \leq c_g n / R \leq 2 \E_g c_g n / R \leq 1/2$ by hypothesis on $c$ and $R$.
    So $\iprod{y_g,v_i} = \Delta_g(g_i \pm \delta_{ig})$ with $|\delta_{ig}| \leq 1/2$, which completes the proof.
\end{proof}

\section{One-Way Communication in the Small-Advantage Regime}
\label{sec:qrac-lb}

In this section we study (quantum) communication complexity of the index function.
We adopt terminology from the quantum information literature, where a one-way protocol for the index function is called a (quantum) random access code.

\begin{definition}[Random Access Code]
    A $(m,n,\epsilon)$ random access code is a map from messages $g \in \{ \pm 1\}^n$ to distributions over $m$-bit codewords $y_g$ together with a family of decoding procedures $D_1,\ldots,D_n \, : \, \{ \pm 1\}^m \rightarrow \{\pm 1\}$ such that for all $g$ and all $i$, the decoding procedure $D_i$ run on input $y_g$ outputs $g_i$ with probability at least $1-\epsilon$ (over the random choice of encoding of $g$).
\end{definition}

\begin{definition}[Quantum random access code]
    A \emph{quantum} random access code is a map from message $g$ to quantum states $\{y_g\}_{g \in \sbits^n}$ together with a family of quantum decoding procedures $D_1,\ldots,D_n \, : \, \{ \pm 1\}^m \rightarrow \{\pm 1\}$.
    We call the code \emph{pure} if $y_g$ is a pure state; otherwise it may be a mixed state.
    The decoding procedure $D_i$ have the property that for all $g$ and all $i$, $D_i$ run on input $y_g$ outputs $g_i$ with probability at least $1-\epsilon$, where the probability is over any classical randomness in the possibly mixed state $y_g$ as well as the randomness in the outcomes of measurements made by $D_i$.
\end{definition}

In the regime that $\epsilon$ is a constant independent of $n,m$, classical information theory shows that $m \geq \Omega(n)$ for classical codes.
This is not so obvious for the quantum case, since the decoding procedure $D_i$ may destroy the state $y_g$ and make other bits unreadable.
Nonetheless, a clever application of the Holevo bound together with an inductive argument shows that $m \geq \Omega(n)$ in quantum case as well \cite{ambainis2002dense}.

For applications to discrepancy, we are interested in the case that $\e$ is close to $1/2$ -- i.e. $\e = 1/2 - \eta$ for some $\eta = \eta(n) \rightarrow 0$.
Of particular interest is the regime $\eta = \Theta(1/\sqrt{n})$ -- this is the most relevant regime for Spencer-style discrepancy bounds, and it also turns out to be interesting from the perspective of random access codes, whose behavior is rather different for $\eta \gg 1/\sqrt{n}$ and $\eta \ll 1/\sqrt{n}$.

To carry out our application to discrepancy, we need lower bounds for somewhat weaker notions of random access codes.

\begin{definition}[Weakness]
    A $(m,n,\epsilon)$ random access code is called $(\delta,t)$-\emph{weak} (for a subset $\cS \subseteq \sbits^n$) if it contains messages $y_g$ only for $g \in \cS$, where $\log \abs{\cS} \geq n-t$, and if for each $g$ at least a $(1-\delta)$ fraction of the decoding procedures $D_i$ yield the bit $g_i$ with probability $1-\epsilon$.
    (So for each $g$ there may be as many as $\delta n$ bad coordinates which are not decoded by their corresponding decoding procedures.)
\end{definition}

We prove the following main result for both classical and pure quantum random access codes.
Adapting the same proof, afterwards we prove Lemma~\ref{lem:rac-lb-matrix-dependent}.

\begin{theorem}[Lower Bound for Low-Signal (Pure) Random Access Codes]
    \label{thm:rac-lb}
  Let $\{y_g\}_{g \in \{ \pm 1\}^n}$ be an $(m,n,\tfrac 12 - \eta)$ classical random access code or pure quantum random access code.
  Then
    \[
        m \geq \log n - \log \log \frac 1 {\eta^2} - O(1)\mper
    \]
    And if $\eta \geq 10/\sqrt{n}$,
    \[
        m \geq \log \frac 1 {\eta^2} + \Omega(\eta^2 n)\mper
    \]
    Furthermore, the same inequalities hold if $\{y_g\}$ is $(\delta,t)$-weak, so long as $\delta \leq \delta_0$ and $t \leq t_0 n$ for some universal constants $\delta_0,t_0$.
\end{theorem}

We conjecture that the same bounds as in Theorem~\ref{thm:rac-lb} hold also for general quantum random access codes, where $y_g$ may be a mixed state -- such bounds would imply the Matrix Spencer conjecture.

The proof of Theorem~\ref{thm:rac-lb} has two parts.
First, we generalize the argument of \cite{ambainis2002dense} to the case of weak quantum random access codes (requiring only a slight adaptation of the arguments of \cite{ambainis2002dense}).
Then, we prove Theorem~\ref{thm:rac-lb} by applying the resulting lemma not to the random access code we start with, but instead to the code we get by amplifying the success probability by repeating the code.

\begin{lemma}[Lower bound for WQRACs, adapted from \cite{ambainis2002dense}]
    \label{lem:wqrac-lb}
    If there exists an $(m,\delta,\epsilon)$-WQRAC for a subset $\cS \subset \sbits^n$ with $\log |\cS| = n - t$, then
    \[ m \geq (1 - H(\delta) - H(\epsilon)) \cdot n - t \mper \] 
\end{lemma}

Now we prove Theorem~\ref{thm:rac-lb} and its refinement Lemma~\ref{lem:rac-lb-matrix-dependent}, deferring the proof of Lemma~\ref{lem:wqrac-lb} to the end of this section.

\begin{proof}[Proof of Theorem~\ref{thm:rac-lb}]
    Let $r > 0$ be an integer.
    Consider the \emph{amplified} random access code which uses
    \begin{itemize}
        \item $r$ independent draws $y_g^1,\ldots,y_g^r$ from the distribution of messages encoding $g$, in the classical case, and
        \item $y_g^{\tensor r}$, that is, $r$ copies of the state $y_g$, in the quantum case,
    \end{itemize}
    as the encoding of $g$.
    To decode the bit $i$, run the decoding procedure $D_i$ on each copy and take a majority vote.
    This gives a $(\delta,t)$-weak random access code for $\cS$ with failure probability $\e' = \Pr( \text{Bin}(r, 1/2 + \eta) < r/2) \leq \exp(-r\eta^2)$.
    Take $r = 1/\eta^2$, so that this is at most $1/e$.

    We claim that in classical case, the message $y_g^1,\ldots,y_g^r$ can be expressed using just
    \begin{align}
        \log \binom{2^m+1/\eta^2}{1/\eta^2} \leq \min \Paren{2^m, \frac 1 {\eta^2}} \cdot \log \Paren{e\cdot \frac{2^m+\frac 1 {\eta^2}}{\min\Paren{2^m, \frac 1 {\eta^2}}}}
        \label{eq:rac-lb-1}
    \end{align}
    bits, and similarly with at most the above number of qubits for the pure quantum case.

    \textbf{Classical: } The majority-vote decoding procedure only needs to know the frequency of each of the $2^m$ possible messages among $y_g^1,\ldots,y_g^{1/\eta^2}$.
    By a ``stars and bars'' argument, the number of such frequency-counts is at most $\binom{2^m+1/\eta^2}{1/\eta^2}$.

    \textbf{Quantum: } The state $y_g^{\tensor 1/\eta^2}$ lies in the \emph{symmetric subspace}, $\Span \{ x^{\tensor 1/\eta^2} \, : \, x \in \R^{2^m} \}$.
    This subspace has dimension at most $\binom{2^m + 1/\eta^2}{1/\eta^2}$ \cite{harrow-church} (by the same stars and bars argument).

    Applying Lemma~\ref{lem:wqrac-lb} to this amplified code gives us the following bound.
    \begin{align}
        \min \Paren{2^m, \frac 1 {\eta^2}} \cdot \log \Paren{e\cdot \frac{2^m+\frac 1 {\eta^2}}{\min\Paren{2^m, \frac 1 {\eta^2}}}} \geq (1 - H(\delta) - H(1/e))n - t \geq \frac n 2 \mper \label{eq:pure-lb-1}
    \end{align}
    where we have used the hypotheses on $\delta$ and $t$.

    Suppose first that we take the bound
    \[
        2^m \log \Paren{e \cdot \frac{2^m + 1/\eta^2}{2^m}} \geq n/2\mper
    \]
    Then we have
    \[
        2^m \cdot \Paren{ \log \frac 1 {\eta^2} + O(1)} \geq \frac n 2\mper
    \]
    Hence,
    \[
        m \geq \log n - \log \log \frac 1 {\eta^2} - O(1)
    \]
    as desired.

    Now, suppose that $\eta \geq 10/\sqrt{n}$.
    Let $c = \log 1/\eta^2 - m$, so that $2^m = 2^{-c}/\eta^2$.
    Suppose $c \geq 0$, so that $2^m \leq 1/\eta^2$.
    Then we have
    \[
        2^m \Paren{ \log \frac 1 {\eta^2 2^m} + 3} \geq \frac n 2\mper
    \]
    By definition of $c$, we get $2^m (c + 3) \geq n/2$, which gives $m \geq \log n + \log(1/(c+3)) - 1$.
    On the other hand, by hypothesis $2^m = 2^{-c}/\eta^2$ so $m \leq \log(1/\eta^2) - c$.
    By hypothesis on $\eta$, this means $m \leq \log n - \log(100) - c$.
    This is a contradiction for $c \geq 0$; we must have $c < 0$.

    Put differently, we may assume that $1/\eta^2 \leq 2^m$.
    Using \eqref{eq:pure-lb-1} again, we obtain
    \[
        \frac 1 {\eta^2} \log \Paren{e \cdot \frac{2^m + \frac 1 {\eta^2}}{\frac 1 {\eta^2}}} \geq n/2\mper
    \]
    This gives
    \[
        \frac{2^m + \frac 1 {\eta^2}}{\frac 1 {\eta^2}} \geq 2^{\eta^2 n/2 - 2}
    \]
    and, rearranging:
    \[
        2^m \geq \frac 1 {\eta^2} \cdot 2^{\eta^2 n/2 - 2} - \frac 1 {\eta^2} = \frac 1 {\eta^2} \cdot \Paren{ 2^{\eta^2 n/2 -2 } - 1}
    \]
    Taking logs,
    \[
        m \geq \log \frac 1 {\eta^2} + \log \Paren{2^{\eta^2 n/2 - 2} - 1} \geq \log \frac 1 {\eta^2} + \Omega(\eta^2 n)
    \]
    where we used the assumption $\eta \geq 10/\sqrt{n}$ for the second inequality.
\end{proof}

\begin{proof}[Proof of Lemma~\ref{lem:rac-lb-matrix-dependent}]
    The proof is identical to that of Theorem~\ref{thm:rac-lb}, except that we construct the decoding procedures for the amplified code as follows.

    \textbf{Decoding the amplified code: } For each $i$, the matrix $A_i$ induces the following measurement procedure: measure in the eigenbasis $a_1,\ldots,a_n$ of $A_i$, and on receiving outcome $j$, output the eigenvalue of $A_i$ associated to $a_j$.
    Given $r$ copies of the state $y_g$, to decode the $i$-th bit, run the aforementioned measurement procedure on each copy of $y_g$ and average the results.
    Output $1$ if the sum is positive and $-1$ otherwise.

    \textbf{Analysis of decoding:} We claim that for some choice of $r = O(\|\E_{i \sim [n]} A_i^2\| / \E_g \eta_g^2)$, the above decoding procedure yields a weak random access code for a subset $\cS' \subseteq \cS$ of size $|\cS'| \geq (1-\delta) |\cS|$, with failure probability at most $1/e$.
    After this is established, the proof can proceed as in Theorem~\ref{thm:rac-lb}, with $\eta = \Theta(\sqrt{\E_g \eta_g^2 / \| \E_{i \sim [n]} A_i^2 \|})$.

    In decoding the $i$-th bit, the decoding procedure produces a sum of i.i.d. random variables $X_1,\ldots,X_r$, each with mean $\eta_g g_i$ and variance $\iprod{y_gy_g^\top, A_i^2}$.
    The average $X = \sum_{i=1}^r X_i / r$ has $\E X = \eta_g g_i$ and variance $\Var(X) = \iprod{y_g y_g^\top, A_i^2} / r$.
    By Chebyshev's inequality, for fixed $i$ and $g$, the probability of incorrectly decoding the bit $g_i$ is at most
    \[
        \Pr \Paren{|X - \E X| > \eta_g} \leq \frac{\eta_g^2}{\iprod{y_g y_g^\top, A_i^2}/r}\mper
    \]
    This is at most $1/e$ so long as $r \geq e \cdot \iprod{y_g y_g^\top, A_i^2} / \eta_g^2$.
    By Markov's inequality, for any $\delta > 0$, if we choose $r \geq C(\delta) \|\E_{i \sim [n]} A_i^2 \| / \E_g \eta_g^2$, we will have $r \geq e \cdot \iprod{y_g y_g^\top, A_i^2} / \eta_g^2$ for at least a $(1-\delta)$ fraction of pairs $(g,i)$ for which the assumptions of the theorem apply.
\end{proof}

\subsection{Proof of Lemma~\ref{lem:wqrac-lb}}
To prove Lemma~\ref{lem:wqrac-lb}, we use the entropy coalescence lemma of \cite{ambainis2002dense}, which is itself a corollary of the Holevo bound.

\begin{lemma}[Entropy Coalescence Lemma, \cite{ambainis2002dense}] \label{lem:entropyCoal}
    Let $\rho_{-1}$ and $\rho_1$ be two density matrices and let $\rho = (1-\beta) \cdot \rho_{-1} + \beta \cdot \rho_1$ be their mixture for some $\beta \in (0,1)$.
    If there is a measurement with outcome $-1$ or $1$ such that making the measurement on $\rho_b$ yields the bit $b$ with probability at least $p$, then
    \[
        S(\rho) \geq (1-\beta) \cdot S(\rho_{-1}) + \beta \cdot  S(\rho_1)) + (H(\beta) - H(p))
    \]
    where $H$ is the binary entropy function and $S$ is the von Neumann entropy. 
\end{lemma}

Now we can prove Lemma~\ref{lem:wqrac-lb} essentially by following the argument of \cite{ambainis2002dense} while throwing out the $\e$-fraction of $g$'s where $Y_g$ is not decodeable to $g$.

\begin{proof}[Proof of Lemma~\ref{lem:wqrac-lb}]
    For each $g \in \cS$, there is a set of at most $\delta \cdot n$ {\it bad} coordinates $i \in [n]$ for which the corresponding decoding procedures $D_i$ do not produce the bit $g_i$ with probability $1-\epsilon$.  Let $I_g \in \binom{n}{\delta n}$ denote this subset.
    Let $I \in \binom{n}{\delta n}$ denote the subset that is the bad set for the largest number of strings $g \in \cS$, and let 
    \[ \cS_I = \{ g \in \cS| I_g = I\} \] 
    Clearly, $\abs{ \cS_I } \geq \abs{\cS}/ \binom{n}{\delta n}$.

 	Without loss of generality, let us assume that the set $I$ consists of the last $\delta n$ coordinates, i.e., $I = \{(1-\delta)n, \ldots, n\}$.
    Let $\Theta$ denote the uniform distribution over  $g \in \cS_I$.
    Let $g \in \sbits^n$ denote a random sample from the distribution $\Theta$.
    For every $L < (1-\delta) \cdot n$, let $\Theta_L$ denote the marginal distribution of $\Theta$ over the first $L$ coordinates.
    For every $\alpha \in \sbits^L$, let us define
    \[ \rho_{\alpha} = \E_{g \sim \Theta}[Y_g | g_L = \alpha] \]

    By definition, we can write
    \[ \rho_{\alpha} = \Pr[g_{L+1} = +1| g_L = \alpha] \cdot \rho_{\alpha, +1} + \Pr[g_{L+1} = -1| g_L = \alpha] \cdot \rho_{\alpha, -1} \]

    Applying the entropy coalescence lemma \pref{lem:entropyCoal}, we get that
    \[ S(\rho_{\alpha}) \geq \Pr[g_{L+1} = +1| g_L = \alpha] \cdot S(\rho_{\alpha, +1}) + \Pr[g_{L+1} = -1| g_L = \alpha] \cdot S(\rho_{\alpha, -1})  +  H(\Pr[g_{L+1} = +1| g_L = \alpha]) - H(\epsilon) \]

    Averaging the above inequality over $\alpha$ drawn from the distribution $\Theta_L$,
    \[ \E_{\alpha \sim \Theta_L} [S(\rho_{\alpha})] \geq \E_{\alpha' \sim \Theta_{L+1}}[S(\rho_{\alpha'})] + H(g_{L+1} | g_L) - H(\epsilon) \]

    Summing up the above inequality over $L = 0,\ldots, (1-\delta) n-1$,
    \begin{align*}
     S(\rho_{\emptyset})] & \geq \E_{\alpha \sim \Theta}[S(\rho_{\alpha})] + \sum_{L = 1}^n H(g_{L+1} | g_L) - (1-\delta) n \cdot H(\epsilon)  \\
     					  & \geq \E_{\alpha \sim \Theta}[S(\rho_{\alpha})] + H(g) - (1-\delta) n \cdot H(\epsilon)  \\
     					  & \geq 0 + H(g) - (1-\delta)n \cdot H(\epsilon)\\
     					  & \geq \log |\cS_I| - (1-\delta) n \cdot H(\epsilon)
    \end{align*}
    The lower bound follows by observing that $S(\rho_{\emptyset}) \leq m$ and $\binom{n}{\delta n} \leq 2^{H(\delta) \cdot n}$.
\end{proof}

\section{Sketching}
\label{sec:sketching}

In this section, we will present a random sketch that preserves evaluations of a set of quadratic forms.
Fix a set of symmetric matrices $A_1,\ldots,A_n \in \R^{D \times D}$ and a point $y \in \R^D$.
Consider the linear sketch that samples a random matrix $S \in \R^{d \times D}$ with entries in $\cN(0,1/d)$ and maps,
\[ A_i \in \R^{D \times D} \to  S A_i S^{\top} \in R^{d \times d} \]
and 
\[ y \in \R^D \to Sy \in \R^d \]
We will show that this sketch approximately preserves the value of the quadratic forms $y^{\top} A_i y$ with good probability.
The rest of this section is devoted to bounds on the expectation, variance of the sketched quadratic forms and the spectral norm of sketched matrices.
These guarantees are captured by the following lemma.
%

\begin{lemma}[Main sketching lemma]
    \label{lem:sketch-main}
    Let $A_1,\ldots,A_n \in \R^{D \times D}$ be symmetric with $\Tr A_i = 0$.
    Let $A = \sum_{i=1}^n A_i^2$.
    Let $y \in \R^D$ be a unit vector.
    Finally, let $d > 0$ and let $S \in \R^{d \times D}$ have iid entries from $\cN(0,1/d)$.
    Then the following all hold:
    \begin{enumerate}
        \item \emph{Expectation:} For all $i$, $\E \iprod{(Sy)(Sy)^\top,SA_iS^\top} = \Paren{1 + \frac 1 d} \iprod{yy^\top, A_i}$,
        \item \emph{Variance:} The average variance across $i =1, \ldots, n$ is bounded:
            \[
                \sum_{i=1}^n \Var \left [ \iprod{(Sy)(Sy)^\top, S A_i S^\top} \right ] \leq O \Paren{ \frac{\|A\|}{d} + \frac{\Tr A}{d^2}}\mcom \text{ and}
            \]
        \item \emph{Spectral norm:} The following matrix has bounded spectral norm:
            \[
                \E \Norm{\sum_{i=1}^n (S A_i S^\top)^2 } \leq O \Paren{\|A\| + \frac{\Tr A}{d}}\mper
            \]
    \end{enumerate}
\end{lemma}

The proof of the main sketching lemma may be found across the following three subsections, in Lemmas~\ref{lem:sketch-quadratic-form-expectation},\ref{lem:sketch-quadratic-form-variance}, and in Section~\ref{sec:sketch-spectral-norm}.

\subsection{Expectation}
\begin{lemma}(Expected Value)
  \label{lem:sketch-quadratic-form-expectation}
  Let $y \in \R^D$.
  Let $A \in \R^{D \times D}$ be symmetric.
  Let $S \in \R^{d \times D}$ have i.i.d. entries from $\cN(0,1/d)$.
  Then
  \[
  \E \iprod{(Sy)(Sy)^\top, SA S^\top } = \left( 1 + \frac{1}{d} \right) \iprod{yy^{\top } , A} + \frac{\norm{y}^2 }{d} \Tr A \mper
  \] 
\end{lemma}
\begin{proof}[Proof of Lemma~\ref{lem:sketch-quadratic-form-expectation}]
      Let $A = \sum_{i} \lambda_i a_i a_i^{\top} $ be the eigendecomposition of $A$ with $\norm{a_i} = 1 $. 
      Then, 
      \begin{align*}
        \E \left[ \iprod{(Sy)(Sy)^\top, SA S^\top } \right] &= \E \left[ \iprod{(Sy)(Sy)^\top, S (\sum_i \lambda_i a_i a_i^{\top} ) S^\top } \right] \\ 
        & = \sum_i \lambda_i \E \left[ \iprod{(Sy)(Sy)^\top, S ( a_i a_i^{\top} ) S^\top } \right] \\ 
        & = \sum_i \lambda_i \E \left[ \iprod{ Sy ,Sa_i }^2 \right].
      \end{align*}
      Let us compute each term separately. Denote by $S_j$ the $j$th row of $S$. 
      Then, 
      \begin{align*}
        \E \left[ \iprod{Sy,Sa_i}^2 \right]  =& \E \left[ \left(  \sum_k \iprod{S_k , y } \iprod{S_k ,a_i } \right)^2 \right] \\ 
         =& \E \left[ \sum_{\ell , k} \iprod{S_k,y} \iprod{S_k,a_i} \iprod{S_{\ell},y} \iprod{S_{\ell},a_i }  \right] \\ 
        =& \sum_{\ell , k} \E \left[\iprod{ S_k , y }  \iprod{ S_k ,a_i} \right] \E \left[\iprod{S_\ell , y} \iprod{S_{\ell } , a_i }\right] + \E \left[\iprod{ S_k , y } \iprod{ S_{\ell} ,y} \right] \E \left[\iprod{S_{\ell} , a_i} \iprod{S_{k } , a_i } \right] \\ & + \E \left[ \iprod{ S_k , y }  \iprod{ S_{\ell} ,a_i}\right] \E \left[\iprod{S_\ell , y} \iprod{S_{k } , a_i } \right]         \qquad \using{Wick's theorem}  
      \end{align*}
      Note that $ \E \iprod{ S_k , y }  \iprod{ S_k ,a_i} = \E y S_k^{\top } S_k a_i = \frac{1}{d} \iprod{y,a_i }  $. 
      Also, note that the second and the third terms are zero if $ \ell \neq k $. 
      Thus, we get 
      \begin{equation*}
        \E \left[ \iprod{Sy,Sa_i}^2 \right] = \iprod{y,a_i}^2 + \frac{1}{d} \norm{y}^2 \norm{a_i}^2 + \frac{1}{d} \iprod{y,a_i}^2  
      \end{equation*}
      Summing over all $i$ gives us 
      \begin{align*}
        \sum_i \lambda_i \E \left[ \iprod{Sy,Sa_i}^2 \right] =& \left( 1 + \frac{1}{d} \right) \sum_i \lambda_i \iprod{y,a_i}^2 + \frac{\norm{y}^2 }{d} \sum_i \lambda_i \\ 
        =& \left( 1 + \frac{1}{d} \right) \iprod{yy^{\top } , A} + \frac{\norm{y}^2 }{d} \Tr A 
      \end{align*}
      as required. 
\end{proof}

\subsection{Variance}

\begin{lemma}
  \label{lem:sketch-quadratic-form-variance}
  Let $y \in \R^D$ have $\|y\|^2 = 1$.
  Let $A_1,\ldots,A_n \in \R^{D \times D}$ be symmetric matrices with $\Tr A_i = 0$.
    Let $B = \sum_{i \leq n} A_i^2$.
  Let $S \in \R^{d \times D}$ have i.i.d. entries from $\cN(0,1/d)$.
  Then
  \[
      \sum_{i \leq n} \E_S \iprod{(Sy)(Sy)^\top, SA_i S^\top }^2 - \Paren{\E_S \iprod{(Sy)(Sy)^\top, S A_i S^\top}}^2 \leq O\Paren{\frac{\|B\|}{d} + \frac{\Tr B}{d^2}}\mper
  \] 
\end{lemma}
\begin{proof}
    By using Proposition~\ref{prop:sketch-quadratic-form-variance-1} on each term in the sum and simplifying with $\Tr A_i = 0$ and the bound $\iprod{y,A_i y}^2 \leq \|y\|^2 \iprod{y, A_i^2 y}$, we obtain that the above is at most
    \begin{align*}
        O(1/d) \cdot \Iprod{y, \sum_{i \leq n} A_i^2 y}
        + O(1/d^2) \cdot \sum_{i \leq n} \|A_i\|_F^2\mper
    \end{align*}
    The result follows by observing that $\Tr \sum_{i \leq n} A_i^2 = \sum_{i \leq n} \|A_i\|_F^2$.
\end{proof}

\begin{proposition}
    \label{prop:sketch-quadratic-form-variance-1}
    Let $y \in \R^D$.
    Let $A \in \R^{D \times D}$ be symmetric.
    Let $S \in \R^{d \times D}$ have i.i.d. entries from $\cN(0,1/d)$.
    Then
\begin{align*}
    & \E \iprod{(Sy)(Sy)^\top, S A S^\top}^2 - \Paren{ \E \iprod{(Sy)(Sy)^\top, S A S^\top}}^2\\
    &  = \Theta(1/d)  \iprod{y,Ay}^2 + \Theta(1/d) \|y\|^2 \iprod{y,A^2 y})\\
        & + \|y\|^2 \cdot ( \Theta(1/d^2) \iprod{y,Ay} \Tr A + \Theta(1/d^2) \|y\|^2 \|A\|_F^2 )\\
        & + \Theta(1/d^3) \|y\|^4 (\Tr A)^2
    \end{align*}
\end{proposition}
\begin{proof}
    Let $\lambda_1,\ldots,\lambda_D$ be the eigenvalues of $A$ with associated eigenvectors $a_1,\ldots,a_D$, so that $A = \sum_{i=1}^D \lambda_i a_i a_i^\top$.
    Then we can expand as
    \[
        \sum_{i,j \leq D} \lambda_i \lambda_j \Paren{ \E \iprod{Sy,Sa_i}^2 \iprod{Sy,Sa_j}^2 - \E \iprod{Sy,Sa_i}^2 \E \iprod{Sy,Sa_j}^2}\mper
    \]
    Applying Proposition~\ref{prop:sketch-variance} to each term, we get that the above is equal to
    \begin{align*}
        & \sum_{i,j \leq D} \lambda_i \lambda_j \left(\Theta(1/d) \iprod{y,a_i}^2 \iprod{y,a_j}^2 + \Theta(1/d) \|y\|^2 \iprod{y,a_i} \iprod{y,a_j} \iprod{a_i,a_j}\right) \\
        & + \sum_{i,j \leq D} \lambda_i \lambda_j  \|y\|^2 \left( \Theta(1/d^2) \iprod{y,a_i}^2 \|a_j\|^2 + \Theta(1/d^2) \iprod{y,a_j}^2 \|a_i\|^2 + \Theta(1/d^2) \|y\|^2 \iprod{a_i,a_j}^2\right)\\
        & + \Theta(1/d^3) \cdot \sum_{i,j \leq D} \lambda_i \lambda_j \|y\|^4 \|a_i\|^2 \|a_j\|^2\mper
    \end{align*}
    Since $a_1,\ldots,a_D$ are orthonormal, this simplifies to
    \begin{align*}
        & \Theta(1/d) \iprod{y,Ay}^2 + \Theta(1/d) \cdot \|y\|^2 \iprod{y,A^2 y}\\
        & + \|y\|^2 \cdot ( \Theta(1/d^2) \iprod{y,Ay} \Tr A + \Theta(1/d^2) \|y\|^2 \|A\|_F^2 )
        + \Theta(1/d^3) \|y\|^4 (\Tr A)^2
    \end{align*}
    as desired.
\end{proof}

\begin{proposition}
    \label{prop:sketch-variance}
    Let $y,a,b \in \R^D$.
    Let $S \in \R^{d \times D}$ have i.i.d. entries from $\cN(0,1/d)$.
    Then 
    \begin{align*}
        \E \iprod{Sy,Sa}^2 \iprod {Sy, Sb}^2 - \E \iprod{Sy,Sa}^2 \E \iprod{Sy,Sb}^2 & = \Theta(1/d) \cdot \iprod{y,a}^2 \iprod{y,b}^2 \\
        & + \Theta(1/d) \cdot \|y\|^2 \iprod{y,a}\iprod{y,b}\iprod{a,b} \\
        & + \Theta(1/d^2) \cdot \|y\|^2 \iprod{y,b}^2 \|a\|^2 \\
        & + \Theta(1/d^2) \cdot \|y\|^2 \iprod{y,a}^2 \|b\|^2 \\
        & + \Theta(1/d^2) \cdot \|y\|^4 \iprod{a,b}^2 \\
        & + \Theta(1/d^3) \cdot \|y\|^4 \|a\|^2 \|b\|^2\mper
    \end{align*}
\end{proposition}
\begin{proof}
    Let $S_i$ be the $i$-th row of $S$, for $i \in [d]$.
    We can expand the above as
    \begin{align}
        \label{eq:sketch-1}
        \sum_{i,j,k,\ell \in [d]} & \E \iprod{S_i,y}\iprod{S_i,a} \iprod{S_j,y}\iprod{S_j,a} \iprod{S_k,y}\iprod{S_k,b} \iprod{S_\ell,y} \iprod{S_\ell,b} \\
        & - \E \iprod{S_i,y}\iprod{S_i,a} \iprod{S_j,y}\iprod{S_j,a}  \cdot \E \iprod{S_k,y}\iprod{S_k,b} \iprod{S_\ell,y} \iprod{S_\ell,b} 
    \end{align}
    Each term in the sum \eqref{eq:sketch-1} above expands in terms of perfect matchings on the following labeled $8$-vertex graph:

    \begin{figure}[h!]
    \begin{center}
    \includegraphics[scale=0.15]{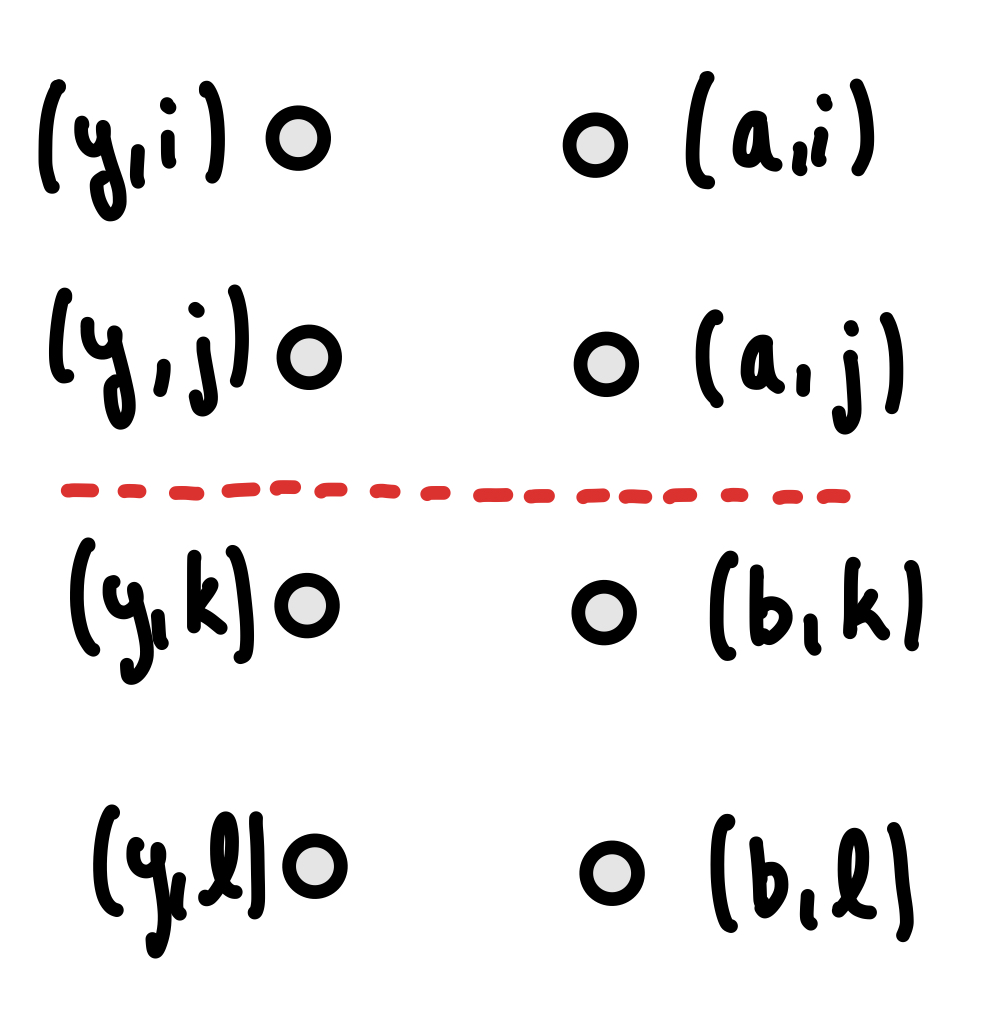}
    \end{center}
    \caption{The graph $G$}
    \end{figure}

    Concretely, by Wick's theorem, the sum \eqref{eq:sketch-1} is equal to the output of the following algorithm, for some functions $d_\alpha = d_\alpha(d) = \Theta(d^{-\alpha})$ with $\alpha \in \{0,1,2,3\}$.

    \begin{enumerate}
        \item Let $\text{output} = 0$.
        \item For every perfect matching $M$ in the graph $G$ such that $M$ contains at least one edge crossing the red cut in $G$:
            \begin{enumerate}
                \item Let $\beta$ be the number of connected components in the (multi)graph that $M$ induces on vertices $\{ i,j,k,\ell \}$ (where e.g. vertices $(y,i),(a,i)$ are collapsed to a single vertex). Let $\alpha = 4 - \beta$
                \item Let
                    \[
                        \text{output} = \text{output} + d_\alpha \cdot \prod_{ \{(v,s),(w,t)\} \in M } \iprod{v,w}\mper
                    \]
            \end{enumerate}
        \item Output $\text{output}$.
    \end{enumerate}
    The terms accumulated in $\text{output}$ are all monomials in the following variables:
    \[
        d_0, d_1, d_2, d_3, \|y\|^2, \|a\|^2, \|b\|^2, \iprod{y,a}, \iprod{y,b}, \iprod{a,b}\mper
    \]
    We need to find the leading-order coefficient (i.e. $d_0,d_1,d_2$ or $d_3$) on each of the following monomials.
    \begin{align*}
        & \|y\|^4 \|a\|^2 \|b\|^2,\\
        & \|y\|^4 \iprod{a,b}^2,\\
        & \|y\|^2 \iprod{y,a}^2 \|b\|^2,\\
        & \|y\|^2 \iprod{y,b}^2 \|a\|^2,\\
        & \|y\|^2 \iprod{y,a}\iprod{y,b} \iprod{a,b},\\
        & \iprod{y,a}^2\iprod{y,b}^2
    \end{align*}
    (One can easily check by hand that this list accounts for all possible matchings in the graph $G$ above.)
    We proceed by cases.
    \begin{enumerate}
        \item $\|y\|^4 \|a\|^2 \|b\|^2$. For a matching $M$ to produce this term and have some edge cross the red cut, $M$ must match $\{(y,i),(y,j)\}$ to $\{(y,k),(y,\ell)\}$.
            And, it must match $(a,i)$ to $(a,j)$ and $(b,k)$ to $(b,\ell)$.
            So the induced graph on $\{i,j,k,\ell\}$ will have just one connected component, and the leading coefficient must be $O(d_3)$.
        \item $\|y\|^4 \iprod{a,b}^2$. WLOG we may assume a matching $M$ producing this term matches $(a,i)$ to $(b,k)$ and $(a,j)$ to $(b,\ell)$.
            Then to create the greatest possible number of connected components in the induced graph on $\{i,j,k,\ell\}$, $M$ must match $(y,i)$ to $(y,k)$ and $(y,j)$ to $(y,\ell)$.
            The induced graph will have $2$ connected components, so the leading coefficient must be $O(d_2)$.
        \item $\|y\|^2 \iprod{y,a}^2 \|b\|^2$. A matching $M$ producing this term must match $(b,k)$ to $(b,\ell)$.
            $M$ must also match both $(y,k),(y,\ell)$ to vertices on the top of the red cut.
            To create the most connected components in the induced graph, $M$ should match $(y,k), (y,\ell)$ either both to ``$i$'' vertices or both to ``$j$'' vertices.
            WLOG suppose it is the latter.
            Then there are two connected components in the induced graph, $\{i\},\{j,k,\ell\}$.
            So the leading coefficient is $O(d_2)$.
        \item $\|y\|^2 \iprod{y,b}^2 \|a\|^2$. Same as $\|y\|^2 \iprod{y,a}^2 \|b\|^2$, by symmetry. Leading coefficient is $O(d_2)$.
        \item $\|y\|^2 \iprod{y,a}\iprod{y,b} \iprod{a,b}$. A matching $M$ producing this term must match some $a$ vertex to a $b$ vertex; WLOG suppose $M$ matches $(a,j)$ to $(b,k)$.
            Then to create the most connected components in the induced graph, $M$ should match $(y,j)$ to $(y,k)$ and $(y,i)$ to $(a,i)$ and $(y,\ell)$ to $(b,\ell)$.
            This gives $3$ connected components, $\{i\},\{\ell\},\{j,k\}$, for a leading coefficient $O(d_1)$.
        \item $\iprod{y,a}^2 \iprod{y,b}^2$. WLOG $M$ matches $(y,j)$ to $(b,k)$ and $(y,k)$ to $(a,j)$, to cross the red cut. Then it can match $(y,i)$ to $(a,i)$ and $(y,\ell)$ to $(b,\ell)$.
            This gives $3$ connected components, for a leading coefficient $O(d_1)$.
    \end{enumerate}
    Thus, we obtain
    \begin{align*}
        \E \iprod{Sy,Sa}^2 \iprod{Sy,Sb}^2 - \E\iprod{Sy,Sa}^2 \E \iprod{Sy,Sb}^2 & = \Theta(1/d) \cdot \iprod{y,a}^2 \iprod{y,b}^2 \\
        & + \Theta(1/d) \cdot \|y\|^2 \iprod{y,a}\iprod{y,b}\iprod{a,b} \\
        & + \Theta(1/d^2) \cdot \|y\|^2 \iprod{y,b}^2 \|a\|^2 \\
        & + \Theta(1/d^2) \cdot \|y\|^2 \iprod{y,a}^2 \|b\|^2 \\
        & + \Theta(1/d^2) \cdot \|y\|^4 \iprod{a,b}^2 \\
        & + \Theta(1/d^3) \cdot \|y\|^4 \|a\|^2 \|b\|^2
    \end{align*}
    as desired.

\end{proof}

\subsection{Spectral norm}
\label{sec:sketch-spectral-norm}

The third claim of Lemma~\ref{lem:sketch-main} follows from the next three lemmas.
The first uses standard decoupling techniques to bound $\| \sum_{i \leq n } (S A_i S)^2 \|$, the spectral norm of a matrix which is a degree-$4$ polynomial in Gaussian variables, in terms of spectral norms of matrices which are degree-$2$ polynomials in Gaussian variables.

\begin{lemma}
    \label{lem:decouple}
For matrices $S, T \in \R^{ d \times D}$ with entries from $N(0,1/d)$ and for every family of symmetric matrices $\{A_1,\ldots,A_n \} \in \R^{D \times D}$,
\[ \E_S \Norm{ \sum_i S A_i S^{\top} S A_i S^{\top} }  \leq \frac{1}{2} \E_{S,T} \Norm{ \sum_i S A_i S^{\top} T A_i T^{\top} + T A_i T^{\top} S A_i S^{\top} } + 2 \E_{S,T}\Norm{ \sum_i S A_i T^{\top} T A_i S^{\top}}\]
\end{lemma}

The next two lemmas bound the terms on the right-hand side of Lemma~\ref{lem:decouple}, starting with the right-most.

\begin{lemma}
    \label{lem:sketch-spectral-1}
  In the setting of Lemma~\ref{lem:decouple},
  \[
    \E \Norm{ \sum_{i \leq n} S A_i T^\top T A_i S^\top } \leq O\Paren{ \Norm{\sum_{i \leq n} A_i^2} + \frac 1 d \cdot \Tr \sum_{i \leq n} A_i^2 }\mper
  \]
\end{lemma}

Finally, we bound the remaining term.

\begin{lemma}
    \label{lem:sketch-spectral-2}
    In the setting of Lemma~\ref{lem:decouple}, if $\Tr A_i = 0$ for all $i$, then
    \[
        \E \Norm{\sum_{i \leq n} S A_i S^\top T A_i T^\top} \leq 0.01 \E \Norm{\sum_{i \leq n} (T A_i T^\top)^2} + O\Paren{\Norm{\sum_{i \leq n} A_i^2} + \frac 1 d \Tr \sum_{i \leq n} A_i^2 }\mper
    \]
\end{lemma}

The last claim of Lemma~\ref{lem:sketch-main} follows by combining Lemmas~\ref{lem:decouple}, \ref{lem:sketch-spectral-1}, and \ref{lem:sketch-spectral-2}, which we now prove in turn.

\subsubsection{Proof of Lemma~\ref{lem:decouple}}

\begin{proof}[Proof of Lemma~\ref{lem:decouple}]
Let us fix $P = \frac{1}{\sqrt{2}} (S+T)$ and $Q = \frac{1}{\sqrt{2}} (S-T)$.  Notice that $P$ and $Q$ have the same law as $S,T$.  Therefore,
\begin{equation}\label{eq:decouple1}
  \E \Norm{ \sum_i S A_i S^{\top} S A_i S^{\top} }  = \E \Norm{ \sum_i P A_i P^{\top} P A_i P^{\top} } 
\end{equation}

Since $\sum_i P A_i P^{\top} P A_i P^{\top}$ and $\sum_i Q A_i Q^{\top} Q A_i Q^{\top}$ are both positive semidefinite matrices, we get 
\begin{equation} \label{eq:decouple2}
   \Norm{ \sum_i P A_i P^{\top} P A_i P^{\top} }  \leq \Norm{ \sum_i (P A_i P^{\top} P A_i P^{\top} + Q A_i Q^{\top} Q A_i Q^{\top})  } 
\end{equation}

For every $i \in [n]$, we can expand out $P A_i P^{\top} P A_i P^{\top} + Q A_i Q^{\top} Q A_i Q^{\top}$ in terms of $S, T$.  All terms that involve an odd number of $T$'s cancel out and we are left with the following identity.

\begin{align}
4 P A_i P^{\top} P A_i P^{\top} + 4 Q A_i Q^{\top} Q A_i Q^{\top}  = & S A_i S^{\top} S A_i S^{\top} + T A_i T^{\top} T A_i T^{\top} \\
                                                                 & + S A_i S^{\top} T A_i T^{\top} + T A_i T^{\top} S A_i S^{\top} \\
                                                                 & + S A_i T^{\top} S A_i T^{\top} + T A_i S^{\top} T A_i S^{\top} \label{eq:term2} \\
                                                                & + S A_i T^{\top} T A_i S^{\top} + T A_i S^{\top} S A_i T^{\top} 
\end{align}
    Using \pref{fact:Mcauchy} (below) with $M = S A_i T^{\top}$, we get that the term in \eqref{eq:term2} is upper bounded in the psd ordering as,
\begin{align}
S A_i T^{\top} S A_i T^{\top} + T A_i S^{\top} T A_i S^{\top} \preceq S A_i T^{\top} T A_i S^{\top} + T A_i S^{\top} S A_i T^{\top} \label{eq:term2rewrite}
\end{align}
Therefore for every $i \in [n]$,
\begin{align}
4 P A_i P^{\top} P A_i P^{\top} + 4 Q A_i Q^{\top} Q A_i Q^{\top}  \preceq & S A_i S^{\top} S A_i S^{\top} + T A_i T^{\top} T A_i T^{\top} \nonumber \\
                                                                 & + S A_i S^{\top} T A_i T^{\top} + T A_i T^{\top} S A_i S^{\top} \nonumber \\
                                                                 & + 2 S A_i T^{\top} T A_i S^{\top} + 2 T A_i S^{\top} S A_i T^{\top} \nonumber
\end{align}
Summing up over all $i \in [n]$, observing that $PA_i P^{\top}P A_i P^{\top}$, $QA_i Q^{\top} Q A_i Q^{\top}$ are positive semidefinite and using the triangle inequality on $\| \|$,
\begin{align*}
4 \E \Norm{ \sum_i (P A_i P^{\top} P A_i P^{\top} + Q A_i Q^{\top} Q A_i Q^{\top})  } \leq & \Norm{ \sum_i S A_i S^{\top} S A_i S^{\top} } + \Norm{ \sum_i T A_i T^{\top} T A_i T^{\top} } \\
                                                                 & + \Norm{ \sum_i S A_i S^{\top} T A_i T^{\top} + T A_i T^{\top} S A_i S^{\top}} \\
                                                                 & + 2 \Norm{ \sum_i S A_i T^{\top} T A_i S^{\top}} + 2 \Norm{ \sum_i T A_i S^{\top} S A_i T^{\top}}
\end{align*}
Finally, taking expectation over $S,T$ and observing that $S,T,P$ have the same distribution,
\begin{align*}
4 \E \Norm{ \sum_i (P A_i P^{\top} P A_i P^{\top} + Q A_i Q^{\top} Q A_i Q^{\top})  } \leq & 2 \E_S \Norm{ \sum_i S A_i S^{\top} S A_i S^{\top} } \\
                                                                 & + \E_{S,T} \Norm{ \sum_i S A_i S^{\top} T A_i T^{\top} + T A_i T^{\top} S A_i S^{\top}} \\
                                                                 & + 4 \E_{S,T} \Norm{ \sum_i S A_i T^{\top} T A_i S^{\top}} 
\end{align*}
The result follows by using the above inequality with \eqref{eq:decouple1} and \eqref{eq:decouple2}.
\end{proof}

\begin{fact} \label{fact:Mcauchy}
For any matrix $M$,
\[ M \cdot M + M^{\top} \cdot {M^{\top}} \preceq MM^{\top} + M^{\top} M  \]
\end{fact}
\begin{proof}
Follows immediately from the identity, 
\[   M  M + M^{\top}  M^{\top} = M  M^{\top} + M^{\top} M - (M-M^{\top})  (M - M^{\top})^{\top} \qedhere   \]
\end{proof}

\subsubsection{Proofs of Lemma~\ref{lem:sketch-spectral-1} and \ref{lem:sketch-spectral-2}}

For both Lemmas we will use the following helpful propositions.

\begin{proposition}\label{prop:sketch-fixed-matrix}
    Let $M \in \R^{D \times D}$ be any matrix and let $S \sim \cN(0,1/d)^{d \times D}$ be a sketching matrix.
    Then
    \[
        \E \|S M S^\top \| \leq O\Paren{\|M\| + \frac{\|M\|_1}{d}}\mper
    \]
\end{proposition}
\begin{proof}
    Let $\cS$ be a $1/16$-th net of the unit sphere in $\R^d$.
    Standard reasoning shows that it suffices to show that
    \[
        \E \|S M S^\top \| \leq \E \max_{w,v \in \cS} \iprod{w,SMS^\top v}\mper
    \]
    Fix $v,w \in \cS$.
    The expression $\iprod{w, SMS^\top v}$ is a degree-2 polynomial in Gaussian variables $S$.
    If $s \in \R^{dD}$ is a vector flattening of $s$, then we can write it as
    \[
        \iprod{w, SMS^\top v} = \iprod{ss^\top, vw^\top \tensor M}\mper
    \]
    We have
    \[
        \E \iprod{ss^\top, vw^\top \tensor M} = \frac 1 d \cdot \Tr (vw^\top \tensor M) \leq \frac 1 d \|M\|_1\mper
    \]
    By the Hanson-Wright inequality, for some universal $c > 0$,
    \[
        \Pr_S \Paren{\iprod{ss^\top, vw^\top \tensor M} \geq \frac 1 d (\|M\|_1 + t)} \leq \exp \Paren{\frac{-ct^2}{\|vw^\top \tensor M\|_F^2 + t \|vw^\top \tensor M\|}}\mper
    \]
    As $\|vw^\top \tensor M\|_F^2 \leq \|M\|_F^2$ and $\|vw^\top \tensor M\| \leq \|M\|$, by a union bound we have
    \[
        \Pr\Paren{ \max_{v,w \in \cS} \iprod{w,SMS^\top v} \geq \frac 1 d \cdot O(\|M\|_1) + t }\leq \exp \Paren{\frac{-ct^2}{\|M\|_F^2 + t \|M\|} + O(d)}\mper
    \]
    Integrating the tail, we find that
    \[
        \E \max_{v,w} \iprod{w, SM^\top S v} \leq \frac 1 d \cdot O(\|M\|_1) + \frac 1 {\sqrt d} \cdot O(\|M\|_F) + O(\|M\|) \leq \frac 1 d \cdot O(\|M\|_1) + O(\|M\|)
    \]
    where the second inequality is Holder's.
\end{proof}

\begin{proposition}
    \label{prop:tensor-cauchy-schwarz}
    Let $A_1,\ldots,A_n \in \R^{D \times D},B_1,\ldots,B_n \in \R^{D' \times D'}$ be symmetric matrices.
    Then
    \[
        \Norm{\sum_{i \leq n} B_i \tensor A_i} \leq \Paren{ \sum_{i \leq n} \|B_i\|_F^2}^{1/2} \cdot \Norm{ \sum_{i \leq n} A_i^2 }^{1/2}\mper
    \]
\end{proposition}
\begin{proof}
    Let $w \in \R^{DD'}$ be a unit vector and let $w_1,\ldots,w_{D'}$ be its rows when viewed as a $D' \times D$ matrix.
    We can expand and use Cauchy-Schwarz:
    \begin{align*}
        w^\top \sum_{i \leq n} (B_i \tensor A_i) w & = \sum_{i \leq n} \sum_{a,c \leq D', b,d \leq D} w_{ab} w_{cd} (B_i)_{a,c} (A_i)_{b,d} \\
        & = \sum_{i \leq n} \sum_{a,c \leq D'} (B_i)_{a,c} \iprod{w_a,A_i w_c} \\
        & \leq \Paren{\sum_{i \leq n} \sum_{a,c \leq D'} (B_i)_{a,c}^2 }^{1/2} \cdot \Paren{ \sum_{i \leq n} \sum_{a,c \leq D'} \iprod{w_a, A_i w_c}^2}^{1/2} \\
        & \leq  \Paren{\sum_{i \leq n} \sum_{a,c \leq D'} (B_i)_{a,c}^2 }^{1/2} \cdot \Paren{ \sum_{i \leq n} \sum_{a,c \leq D'} \|w_a\|^2 \|A_i w_c\|^2 }^{1/2} \\
        & = \Paren{\sum_{i \leq n} \|B_i\|_F^2 }^{1/2} \cdot \Paren{ \sum_{i \leq n} \|w\|^2 \sum_{c \leq D} \|A_i w_c\|^2 }^{1/2} \\
        & \leq \Paren{\sum_{i \leq n} \|B_i\|_F^2 }^{1/2} \cdot \Paren{  \|w\|^2 \sum_{c \leq D} \|w_c\|^2 \Norm{\sum_{i \leq n} A_i^2 }}^{1/2} \\
        & = \Paren{\sum_{i \leq n} \|B_i\|_F^2 }^{1/2} \cdot \Norm{\sum_{i \leq n} A_i^2}^{1/2}\mper \qedhere
    \end{align*}
\end{proof}

Now we can prove Lemmas~\ref{lem:sketch-spectral-1} and~\ref{lem:sketch-spectral-2}.

\begin{proof}[Proof of Lemma~\ref{lem:sketch-spectral-1}]
    Let $M = \sum_{i \leq n} A_i T^\top T A_i$.
    By Proposition~\ref{prop:sketch-fixed-matrix}, for any $T$,
    \[
        \E_S \Norm {\sum_{i \leq n} S A_i T^\top T A_i S^\top } \leq \frac 1 d \cdot O\Paren{ \Tr M  + d \|M\|} \mper
    \]
    By simple calculation, $\E_T \Tr M = \Tr \sum_{i \leq n } A_i^2$.

    To bound $\E_T \|M\|$, let us observe that $M = BB^\top$ where $B \in \R^{nD \times D}$ is the concatenation of $A_1 T^\top ,\ldots,A_n T^\top$.
    So $\|M\| = \|B^\top B\| = \|T \sum_{i \leq n} A_i^2 T^\top \|$.
    Applying Proposition~\ref{prop:sketch-variance} again, we obtain 
    \[
        \E_T \|M\| \leq \frac 1 d O \Paren{\Tr \sum_{i \leq n} A_i^2 + d \cdot \Norm{\sum_{i \leq n} A_i^2}} \mper
    \]
    This finishes the proof.
\end{proof}

\begin{proof}[Proof of Lemma~\ref{lem:sketch-spectral-2}]
    Let $\cS$ be a $1/16$-th net of the unit sphere in $\R^d$.
    It will suffice to bound
    \[
        \E_{S,T} \max_{v,w \in \cS} \iprod{w, \sum_{i \leq n} S A_i S^\top T A_i T^\top v}\mper
    \]
    Fix any choice of $T$ and let $s \in \R^{dD}$ be a vector flattening of $S$.
    And, fix $v,w \in \cS$.
    Then
    \[
        \iprod{w, \sum_{i \leq n} S A_i S^\top T A_i T^\top v} = \iprod{ss^\top, \sum_{i \leq n} TA_iT^\top vw^\top \tensor A_i}\mper
    \]
    In expectation over $S$, we have
    \begin{align*}
        \E_S \iprod{w, \sum_{i \leq n} S A_i S^\top T A_i T^\top v} & = \iprod{ss^\top, \sum_{i \leq n} TA_iT^\top vw^\top \tensor A_i}\\
        & = \frac 1 d \sum_{i \leq n} \Tr A_i \cdot w^\top T A_i T^\top v\\
        & = 0
    \end{align*}
    by hypothesis on $\Tr A_1,\ldots,\Tr A_n$.

    By the Hanson-Wright inequality,
    \begin{align*}
        & \E_S \max_{v,w} \iprod{ss^\top, \sum_{i \leq n} T A_i T^\top vw^\top \tensor A_i} \leq O \Paren{ \Norm{\sum_{i \leq n} T A_i T^\top vw^\top  \tensor A_i } + \frac 1 {\sqrt d} \Norm{\sum_{i \leq n} T A_i T^\top vw^\top \tensor A_i }_F }  \mper
    \end{align*}
    We claim that for any constant $c > 0$ we like,
    \begin{align}\label{eq:sketch-spectral-2-1}
        \E_T \Norm{\sum_{i \leq n} T A_i T^\top vw^\top  \tensor A_i }  \leq c \E_T \Norm{\sum_{i \leq n} (T A_i T^\top)^2 } + \frac 1 c \cdot \Norm{\sum_{i \leq n} A_i^2}
    \end{align}
    and
    \begin{align}
        \label{eq:sketch-spectral-2-2}
        \E_T \Norm{\sum_{i \leq n} T A_i T^\top vw^\top  \tensor A_i }_F  \leq  c  \E_T \Norm{\sum_{i \leq n} (T A_i T^\top)^2 } + \frac 1 c \cdot \frac{ \Tr \sum_{i \leq n} A_i^2}{d}\mper
    \end{align}
    which will the proof.

    For \eqref{eq:sketch-spectral-2-1}, we have by Proposition~\ref{prop:tensor-cauchy-schwarz} that
    \begin{align*}
        \Norm{\sum_{i \leq n} T A_i T^\top vw^\top  \tensor A_i } & \leq \Paren{\sum_{i \leq n} \|T A_i T^\top vw^\top \|_F^2 }^{1/2} \cdot \Norm{\sum_{i \leq n} A_i^2}^{1/2}\\
        & \leq \Paren{\sum_{i \leq n} \|T A_i T^\top v\|_2^2 }^{1/2} \cdot \Norm{\sum_{i \leq n} A_i^2}^{1/2}\\
        & \leq \Norm{\sum_{i \leq n} (T A_i T^\top)^2}^{1/2} \cdot \Norm{\sum_{i \leq n} A_i^2}^{1/2}\\
        & \leq c \Norm{\sum_{i \leq n} (T A_i T^\top)^2} + \frac 1 c \Norm{\sum_{i \leq n} A_i^2}\mper
    \end{align*}

    For \eqref{eq:sketch-spectral-2-2}, we have
    \begin{align*}
        \Norm{\sum_{i \leq n} TA_iT^\top vw^\top \tensor A_i}_F^2
        & = \sum_{i,j \leq n} \iprod{T A_i T^\top vw^\top, T A_i T^\top vw^\top} \cdot \iprod{A_i,A_j} \\
        & \leq \Paren{\sum_{i,j \leq n} \|T A_i T^\top vw^\top\|_F^2 \|T A_j T^\top vw^\top\|_F^2 }^{1/2} \Paren{\sum_{i,j \leq n} \|A_i\|_F^2 \|A_j\|_F^2}^{1/2}\\
        & \leq \Norm{\sum_{i \leq n} (T A_i T^\top)^2} \cdot \sum_{i \leq n} \|A_i\|_F^2\mper
    \end{align*}
    Therefore,
    \[
        \frac 1 {\sqrt d} \Norm{\sum_{i \leq n} T A_i T^\top vw^\top  \tensor A_i }_F \leq c \Norm{\sum_{i \leq n} (T A_i T^\top)^2 } + \frac 1 c \cdot \frac 1 d \cdot \Tr \sum_{i \leq n} A_i^2\mper\qedhere
    \]
\end{proof}

\subsection{Purify then Sketch}

With Lemma~\ref{lem:sketch-main} in hand we can prove Lemma~\ref{lem:purify-then-sketch}.
We will need the following fact about the rank of solutions to semidefinite programs. 

\begin{theorem}[Barvinok \cite{barvinok1995problems}, Pataki \cite{pataki1998rank}]
\label{thm:barvinok}
    Any compact spectahedron $\{ Y \, : \, \iprod{Y,A_1} = b_1,\ldots, \iprod{Y,A_m} = b_m, Y \succeq 0\}$ contains $Y$ such that $\rank{Y} \leq 4 \sqrt{m}$.
\end{theorem}

\begin{proof}[Proof of Lemma~\ref{lem:purify-then-sketch}]
    Without loss of generality, by Theorem~\ref{thm:barvinok}, we may assume that $t = \max_g \rank Y_g \leq \min(6\sqrt{n},d)$.
    To produce $y_g$, we use the following algorithm:
    \begin{enumerate}
        \item \emph{Purify:} Let $y_g' \in \R^d \otimes \R^{t}$ be a purification of $Y_g$.
            Let $A_i' = A_i \otimes \Id_{t \times t}$, so that $(y_g')^\top A_i' y_g' = \iprod{Y_g,A_i}$.
        \item \emph{Sketch:} Let $S \in \R^{r \times d t}$ be a sketching matrix with iid entries from $\cN(0,1/r)$.
            Let $y_g = S y_g' / \|S y_g'\|$ and let $B_i = \tfrac d {d+1} S A_i' S^\top$.
    \end{enumerate}
    Now we apply the main sketching lemma~\ref{lem:sketch-main}.
    Noting that $\sum_{i \leq n} (A_i')^2 = \sum_{i \leq n} A_i^2 \tensor \Id_{t \times t} = A \tensor \Id_{t \times t}$, this gives for each $g \in \{ \pm 1\}^n $ and each $i \leq n$,
    \begin{align}
        \E_S \sum_{i=1}^n \Paren{ \iprod{(Sy_g')(Sy_g')^\top, B_i} - \iprod{Y_g, A_i}}^2 \leq O\Paren{\frac{\|A\|}{r} + \frac{t \Tr A}{r^2}}\mper \label{eq:qrac-1}
    \end{align}

    For some constant $C$ we will choose shortly, let us call $g$ \emph{good} if there are at least $(1-\delta)n$ indices $i \in [n]$ such that 
    \[
        \Abs{\iprod{(Sy_g')(Sy_g'), B_i} - \iprod{Y_g,A_i}} \leq C \cdot \Paren{\frac{\|A\|}{nr} + \frac{ t \Tr A}{ n r^2}}^{1/2}\mcom
    \]
    and, additionally, $(1/C) \leq \|Sy_g'\|^2 \leq C$.
    By \eqref{eq:qrac-1}, there is $C = C(\delta)$ such that for each $g$ we have $\Pr(g \text{ is good}) \geq 3/4$.
    Therefore, $\E_S \E_{g \sim \{ \pm 1\}^n } \Ind(g \text{ is good}) \geq 3/4$, and hence there is a choice of $S$ such that $\tfrac 3 4 \cdot 2^n$ $g$'s are good.
    We can obtain the pure state $y_g$ as $Sy_g' / \|S y_g'\|$.
       
    The remaining claim then follows by Markov's inequality applied to $\| \sum_{i \leq n} B_i^2 \|$ (using the bound on $\E \|\sum_{i \leq n} B_i^2\|$ in Lemma~\ref{lem:sketch-main}) and a union bound.
\end{proof}

\section*{Acknowledgements}
AS would like to thank Robert Kleinberg and Ayush Sekhari for enlightening conversations. 
We thank Tselil Schramm, Boaz Barak, Umesh Vazirani, Luca Trevisan, and Raghu Meka for several enlightening conversations as this manuscript was being prepared.
SBH was supported by a UC Berkeley Miller Fellowship and a Simons Postdoctoral Fellowship.

\bibliographystyle{alpha}

\bibliography{discrepancy_refs}

\appendix

\section{Tightness of communication lower bounds}
\label{sec:tightness}

\paragraph{$O(1)$ bits when both players receive random inputs}
We start by sketching a simple $O(1)$-bit classical protocol for the $n$-bit index function where Alice and Bob both receive random inputs.
The players fix random vectors $y_1,\ldots,y_t \in \{ -1,1 \}^n$.
Then, given $x$, Alice sends the index of $y_j$ maximizing $\iprod{x,y_j}$; the maximum value will be around $\sqrt{n \log t}$.
If Bob outputs $y_j(i)$ on input $i$, they achieve
\[
    \E_{x \sim \{ \pm 1\}^n} \E_{i \sim [n]} \Pr(b(a,i) = x_i) \geq \frac 12 + \Omega \Paren{\sqrt{\frac{\log t}{n}}}\mper
\]

\paragraph{$\log n$ bits for worst-case inputs via Hadamard matrices}
Next, we sketch a classical protocol for the $n$-bit index function where Alice sends at most $\log n + 2$ bits and the players have advantage $\Omega(1/\sqrt n)$ over random guessing -- the protocol works even when both Alice and Bob have worst-case inputs.

Without loss of generality we can assume that $n$ is a power of $2$.
We interpret Alice's possible inputs $x \in \{ \pm 1\}^n$ as Boolean functions on $\log n$ bits.
On input $x$, Alice computes the Fourier transform of $x$ as a Boolean function, to obtain $n$ Fourier coefficients $\hat{x}(S)$ for $S \subseteq [\log n]$.
She draws $S$ according to the distribution $\Pr(S) = |\hat{x}(S)| / \sum_{T \subseteq [\log n]} |\hat{x}(T)|$ and sends $S$, using $\log n$ bits, and the sign of $\hat{x}(S)$, using one bit.
Given $i \in [n]$, which we think of as a $\log n$-bit string, Bob outputs $\text{sign}(\hat{x}(S)) \cdot \chi_S(i)$ -- the value of the $S$-th Fourier character on input $i$, with the sign flipped according to the last bit of Alice's message.

For the analysis, fix $x$ and fix $i \in [n]$.
We need to analyze $\Pr(\text{output} = x_i)$, which we can write as
\[
\frac {\sum_S |\hat{x}(S)| \cdot \frac{1 + x_i \cdot \text{sign}(\hat{x}(S))\chi_S(i)}{2} }{\sum_S |\hat{x}(S)|} = \frac 12 - \frac 12 \cdot \frac{\sum_S \hat{x}(S) \cdot x_i \cdot \chi_S(i)}{\sum_S |\hat{x}(S)|}\mper
\]
The expression in the numerator of the last expression is exactly $x_i^2 = 1$.
The denominator satisfies $\sum_S |\hat{x}(S)| \leq \sqrt n \cdot \sqrt{\sum_S |\hat{x}(S)|^2} = \sqrt n$, since $x$ has unit norm as a Boolean function.
So we find that the protocol succeeds with probability at least $\tfrac 12 + \tfrac 1 {2 \sqrt n}$.

\end{document}